\documentclass[journal,10pt]{IEEEtran}
\ifCLASSINFOpdf
\usepackage[pdftex]{graphicx}
\usepackage{amsthm,amsmath,amssymb}
\usepackage{makecell}
\usepackage{graphicx}
\usepackage{mathrsfs}
\usepackage{float}
\usepackage{algorithm}
\usepackage{algpseudocode}
\usepackage{amsmath}
\usepackage{bm}
\usepackage{color}
\usepackage{comment}
\usepackage{makecell}
\usepackage{multirow}
\usepackage{url}
\else
\fi
\usepackage{leftidx}
\usepackage{epstopdf}
\usepackage{amsmath}

\usepackage{subfigure}

\begin{document}

	\title{M$^3$SC: A Generic Dataset for Mixed Multi-Modal (MMM) Sensing and Communication Integration}

	\author{Xiang~Cheng,~\IEEEmembership{Fellow,~IEEE}, Ziwei~Huang,~\IEEEmembership{Graduate~Student~Member,~IEEE}, Lu~Bai,~\IEEEmembership{Member,~IEEE},  Haotian~Zhang,~\IEEEmembership{Graduate~Student~Member,~IEEE}, Mingran~Sun,~\IEEEmembership{Graduate~Student~Member,~IEEE}, Boxun~Liu,~\IEEEmembership{Graduate~Student~Member,~IEEE}, Sijiang~Li,~\IEEEmembership{Graduate~Student~Member,~IEEE}, Jianan~Zhang,~\IEEEmembership{Member,~IEEE}, and Minson~Lee		
		\thanks{X. Cheng, Z. Huang, H. Zhang, M. Sun, B. Liu, S. Li, and J. Zhang are with the State Key Laboratory of Advanced Optical Communication Systems and Networks,
School of Electronics, Peking University, Beijing 100871, P. R. China (email: xiangcheng@pku.edu.cn, ziweihuang@pku.edu.cn, haotianzhang@stu.pku.edu.cn,  mingransun\_pku@163.com, boxunliu\_pku@163.com,  pkulsj@pku.edu.cn, zhangjianan@pku.edu.cn).}
		\thanks{L. Bai is with the Joint SDU-NTU Centre for Artificial Intelligence Research (C-FAIR), Shandong University, Jinan 250100, P. R. China (e-mail: lubai@sdu.edu.cn).}
	\thanks{M. Lee is with Chief Technical Officer, Ever-Florescence Technology, Nanjing 210000, P. R. China
(e-mail: minson@qi-well.com).}}

	\markboth{}
	{Zeng \MakeLowercase{\textit{et al.}}: Bare Demo of IEEEtran.cls for IEEE Journals}
	%


	
		\maketitle
	\vspace{-2cm}
	\begin{abstract}
	The sixth generation (6G) of mobile communication system is witnessing a new paradigm shift, i.e., integrated sensing-communication system.
A comprehensive dataset is a prerequisite for 6G integrated sensing-communication research.
This paper develops a novel simulation dataset, named M$^3$SC,  for mixed multi-modal (MMM) sensing-communication integration, and the generation framework of the M$^3$SC  dataset is further given. To obtain multi-modal sensory data  in  physical space and communication data in electromagnetic space, 
we  utilize AirSim and WaveFarer to  collect multi-modal sensory data and exploit Wireless InSite to  collect communication data. Furthermore, the in-depth integration and precise alignment of AirSim, WaveFarer, and Wireless InSite are achieved.
The M$^3$SC  dataset covers various weather conditions, various frequency bands,  and different times of the day. Currently, the M$^3$SC dataset
 contains 1500 snapshots, including 80 RGB images, 160 depth maps, 80 LiDAR point clouds, 256 sets of mmWave waveforms with 8 radar point clouds, and 72 channel impulse response (CIR) matrices per snapshot, thus totaling 120,000 RGB images, 240,000 depth maps,  120,000 LiDAR point clouds, 384,000 sets of mmWave waveforms with 12,000 radar point clouds, and 108,000 CIR matrices.
The data processing result  presents the multi-modal sensory information and communication channel statistical properties. Finally, the  MMM sensing-communication application, which can be supported by the M$^3$SC dataset, is discussed.

	\end{abstract}
	
	
	\begin{IEEEkeywords}
	6G; multi-modal sensory information;  ray-tracing-based channel information; sensing-communication integration scenarios;  simulation dataset
	\end{IEEEkeywords}

	%
	\IEEEpeerreviewmaketitle

	\section{Introduction}
	%
	%
	%
	
	\IEEEPARstart  Nowadays, the standardization of fifth generation (5G) communications has been completed and 5G networks have been commercially 
launched in 2020 \cite{nature}. 
As the next generation, sixth generation (6G) has attracted tremendous research in  academia and industry \cite{comst-1}--\cite{TWCMY}. Aiming at enriching the spectrum efficiency and reducing the
hardware cost, the 6G communication system is witnessing a major paradigm shift, i.e.,  in-depth  sensing and communication integration \cite{ISAC-1}--\cite{ISAC-IET}. Also, the sensing and communication integration facilitates various 6G applications, e.g., human counting and identification, autonomous driving vehicles, and  area map reconstruction, where widely deployed multi-modal sensors and communication equipment play an important role \cite{ISAC-2}.
Recently, inspired by human synesthesia, an involuntary human neuropsychological trait in which the stimulation of one human sense evokes another human sense, \cite{ISAC-1} refers to these
multi-modal sensors and communication equipment as
``machine senses'', and further introduces the concept of human synesthesia to multi-modal sensing and communication integration. In the context of ``machine senses'', multi-modal sensing-communication integration is referred  as Synesthesia of Machines (SoM). The multi-modal
sensing-communication integration can be achieved in a significantly
intelligent manner via SoM processing, where artificial neural networks (ANNs) serve as the fundamental tool.
To properly support the system design of multi-modal sensing-communication integration, i.e., SoM, a comprehensive dataset containing aligned sensing and communication environmental information is essential. 
Specifically, a comprehensive dataset can provide propagation channel
information in electromagnetic space and  spatial features in physical space,
and is also the lifeblood of  ANNs' learning ability and
the key to the precision of ANNs' output. Meanwhile, such a dataset can support the development of integrated
sensing-communication algorithms,
and further evaluate and compare their performance in a fair manner.

\begin{figure*}[!t]
		\centering	\includegraphics[width=0.99\textwidth]{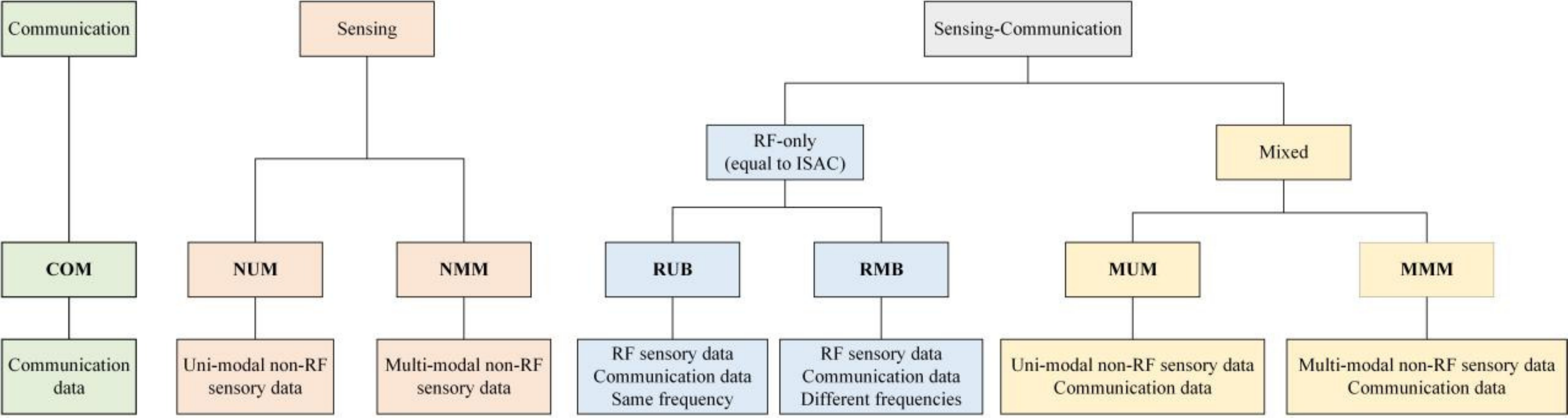}
	\caption{Classification of existing datasets.}
	\label{Classification}
	\end{figure*}
 
Considering the importance of datasets for the system design, extensive datasets have been developed, which can be classified depending on the application, operating frequency, as well as uni-/multi-modal sensor equipment of  system   applicable, as shown in Fig~\ref{Classification}. Datasets for the communication system, named  \emph{communications only (COM)}, solely include  wireless channel data in electromagnetic space. For sensing systems, we divide datasets into \emph{non-radio frequency (non-RF) uni-modal (NUM)} and \emph{non-RF multi-modal (NMM)} datasets.
 The NUM and NMM datasets include uni-modal and multi-modal non-RF sensory data collected in  physical space, respectively.
 To enrich spectrum efficiency, reduce hardware cost, and facilitate more potential applications,  sensing and communications are expected to have an in-depth integration  \cite{ISAC-2}. For integrated sensing-communication systems, we categorize datasets  as the RF-only  dataset and the mixed dataset. The former can support the design of sensing and communication systems operating in  RF, i.e., equal to   integrated sensing and communications (ISAC), and we further divide it into \emph{RF-only uni-band (RUB)} and \emph{RF-only multi-band (RMB)} datasets.  Specifically, the RUB dataset contains the RF sensory data and  the communication data  at the same frequency, which can support the research on dual-functional radar and communications (DFRC). Different from the RUB dataset, the RMB dataset contains the RF sensory data and  the communication data  at different frequencies.
 Finally,
the mixed dataset includes sensory data and communication data collected in non-RF and RF, respectively. Furthermore, we divide the mixed dataset into \emph{mixed uni-modal (MUM)}  and \emph{mixed multi-modal (MMM)}  datasets depending on whether  the dataset contains uni-modal or multi-modal non-RF sensory data.

The authors in \cite{Data-5} developed a measurement dataset, named  KITTI, for NMM sensing systems. The KITTI dataset consists of  calibrated and synchronized RGB images, depth maps, and LiDAR point clouds. To further include the  channel information and support the wireless communication application, a RUB measurement dataset in \cite{WALDO}, named WALDO,  was developed for sensing-communication applications, where sensing and communications were conducted at the  same millimeter wave (mmWave)  frequency under an indoor scenario. The authors in \cite{Data-7} developed
a MMM  measurement dataset, named  DeepSense 6G, for integrated sensing-communication systems. The DeepSense 6G dataset \cite{Data-7}
captured  multi-modal sensory data, such as RGB images and light detection and ranging (LiDAR) point clouds,
and wireless channel data under sub-6 GHz and mmWave  frequency bands. However, the DeepSense 6G dataset \cite{Data-7} ignored
the massive multiple-input multiple-output (MIMO) channel information and snowy weather, and thus cannot support the corresponding integrated sensing-communication application. In addition, although  KITTI, WLADO, and DeepSense  measurement datasets in \cite{Data-5}--\cite{Data-7} can support the validation of fundamental algorithms, it is difficult to flexibly customize desired scenarios owing to the labor and cost concerns.

\begin{table*}[!t]
	\caption{Typical  measurement and simulation datasets.}
	\centering
	\begin{scriptsize}
	\begin{tabular}{cccccccccccc}
		\hline
		\multirow{2}{*}{\textbf{Dataset}} &
		\multirow{2}{*}{\textbf{Type}} &
		\multicolumn{4}{c}{\textbf{Sensory data}} &
		\multicolumn{2}{c}{\textbf{Communication data}} & {\textbf{Weather}}  & \multirow{2}{*}{\textbf{Multi-Scenario}}  & \multirow{2}{*}{\textbf{Time}} &\multirow{2}{*}{\textbf{Source}} 
		\\
		\cline{3-9}  
		
		& & RGB & Depth map & LiDAR & Radar&mmWave& 
		Massive MIMO & Sunny, rainy, snowy  & & &   \\
		\hline
		KITTI \cite{Data-5} & NMM & \checkmark & \checkmark  & \checkmark  & $\times$  &  $\times$ & $\times$ & $\times$ & \checkmark & $\times$ & Measurement\\
		\hline
		WLADO \cite{WALDO} & RUB & $\times$ & $\times$  & $\times$  & \checkmark  &  \checkmark & $\times$ & $\times$ & $\times$ & $\times$ & Measurement\\
		\hline
		DeepSense \cite{Data-7} & MMM & \checkmark & $\times$  & \checkmark   & \checkmark  &  \checkmark & $\times$ & $\times$ & \checkmark & \checkmark & Measurement\\
		\hline
		DeepMIMO \cite{Data-2} & COM & $\times$ & $\times$  & $\times$   & $\times$  &  \checkmark & \checkmark & $\times$ & \checkmark& $\times$ & Simulation\\
		\hline
		LiDARsim \cite{LiDARsim} & NUM & $\times$ & $\times$  & \checkmark   & $\times$  &  $\times$ & $\times$ & \checkmark & \checkmark & $\times$ & Simulation\\
		\hline
		OPV2V \cite{OPV2V} & NMM & \checkmark & \checkmark & \checkmark   & $\times$  &  $\times$ & $\times$ & $\times$ & \checkmark & $\times$ & Simulation\\
		\hline
		SHIFT \cite{SHIFT} & NMM & \checkmark & \checkmark & \checkmark   & $\times$  &  $\times$ & $\times$ & \checkmark & \checkmark & \checkmark & Simulation\\
		\hline
		Radar-COM \cite{Ali} & RMB & $\times$ & $\times$ & $\times$   & \checkmark  &  \checkmark & $\times$ & $\times$ & \checkmark & $\times$ & Simulation\\
		\hline 
		LiDAR-COM \cite{Klautau} & MUM & $\mathbf{\times}$ & $\times$ & \checkmark   & $\times$ &  \checkmark & \checkmark & $\times$ & \checkmark & $\times$ & Simulation\\
		\hline
		ViWi \cite{ViWi} & MMM & \checkmark & \checkmark  & \checkmark    & \textbf{Limited} &  \checkmark & \checkmark & $\times$ & \checkmark & $\times$ & Simulation\\
		\hline
	\end{tabular}
		\end{scriptsize}
	\label{dataset}  
\end{table*}

Due to the limitation of measurement datasets, extensive simulation datasets \cite{Data-2}--\cite{Ali}, \cite{Klautau,ViWi}  are developed as supplements of measurement datasets.
Based upon efficient software with high accuracy,  simulation datasets can achieve an excellent trade-off between complexity and fidelity.  By flexibly adjusting key parameters, the  simulation dataset can further cover diverse application scenarios.
 In \cite{Data-2}, the authors developed a  COM dataset, named  DeepMIMO.
The DeepMIMO dataset \cite{Data-2} intends to promote  machine
learning (ML) applications related to mmWave and massive MIMO based on
  ray-tracing technologies. 
The authors in \cite{LiDARsim} developed a NUM   dataset, named  LiDARsim, which includes the simulated LiDAR point clouds and covers various scenarios. However, the LiDARsim dataset \cite{LiDARsim} is of low information redundancy due to the uni-modal sensor, and thus cannot support applications requiring high sensing robustness.
To address this deficiency, the authors in \cite{OPV2V} developed a NMM dataset, named  OPV2V. The OPV2V dataset \cite{OPV2V} covers  70 vehicular scenarios and consists of RGB images, depth maps, and LiDAR point clouds. To further capture the impact of weather conditions and times of the day, Sun \emph{et al.}  \cite{SHIFT} developed 
a different NMM dataset, named  SHIFT. The SHIFT dataset \cite{SHIFT} has  synthetic RGB images, depth maps, and LiDAR point clouds under diverse vehicular scenarios. Nevertheless, the LiDARsim, OPV2V, and SHIFT datasets in \cite{LiDARsim}--\cite{SHIFT} for sensing systems cannot support integrated sensing-communication applications. To overcome this limitation, by Wireless InSite simulation platform \cite{WI},
Ali \emph{et al.} \cite{Ali} developed a RMB dataset, named Radar-COM in this paper, where communication systems operated at  $73$ GHz band, while   the radar operated at  $76$ GHz band. By employing the non-RF LiDAR simulator in \cite{LIDARS} and  Wireless InSite simulation platform, Klautau \emph{et al.} \cite{Klautau} 
 developed a MUM  dataset, named LiDAR-COM in this paper, consisting of LiDAR point clouds and ray-tracing-based wireless channel data. The LiDAR-COM dataset \cite{Klautau} covered various  scenarios under mmWave and massive MIMO  communications.  
 Although 
the Radar-COM and LiDAR-COM datasets in \cite{Ali} and \cite{Klautau} can promote the development of integrated sensing-communication algorithms, their low  information redundancy and sensing robustness result in reduced accuracy and applicability. 
To overcome this drawback, the authors in \cite{ViWi} proposed 
a MMM dataset,
named  Vision-Wireless (ViWi). 
The ViWi dataset consists of ray-tracing-based wireless channel data collected in Wireless InSite  and multi-modal sensory data, such as RGB images, depth maps, and LiDAR point clouds. Furthermore, the  ray-tracing-based channel data includes sub-6 GHz and mmWave frequency bands and massive MIMO. However,  the ViWi dataset \cite{ViWi} leaves out the simulation of different weather conditions and times of the day, making it impossible to be applied in research on integrated sensing-communication related to weather and time of the day. 

In summary, since sensing-communication integration facilitates diverse applications,  the dataset solely for sensing or communication systems is of low applicability. 
Furthermore, the impact of harsh weather conditions on  communication and sensing systems is a crucial research topic as rainy and snowy days can damage the sensing robustness and communication reliability.  Consequently,   the dataset lacking various weather conditions has limited applications. Meanwhile, due to the respective effects of  
times of the day and frequency bands on the sensing and communication performance, the dataset without different times of the day and various frequency bands is incomplete. Table~\ref{dataset} lists some typical existing datasets.
Currently, a comprehensive MMM simulation dataset, which  covers various weather conditions, frequency bands, and times of the day, is still lacking.

In this paper, we propose M$^3$SC, a framework of developing simulation MMM datasets for multi-modal sensing-communication integration, i.e., SoM. Various
weather conditions, various frequency bands, and
different times of the day are further taken into account.
Currently, there is no software tailored for constructing the MMM dataset for integrated sensing-communication research. However, a comprehensive MMM dataset is a prerequisite for 6G integrated sensing-communication research. To fill this gap, 
we utilize AirSim  \cite{AirSim} and WaveFarer \cite{WaveFarer} to extract spatial features in physical space
 and Wireless InSite \cite{WI} to capture channel information in electromagnetic space. Note that it is not a simple combination of AirSim, WaveFarer, and Wireless InSite. There are two huge challenges, one is the \emph{in-depth integration} of sensing and communications, and the other is the \emph{precise alignment} of  physical
space and  electromagnetic space.
For the former challenge, 
to achieve the in-depth integration of sensing and communications, AirSim and WaveFarer need to meet the communication requirements, such as collecting space-air-ground-sea environmental information. Nonetheless, AirSim and WaveFarer cannot naturally meet this requirement. Meanwhile, Wireless InSite needs to meet sensing requirements, such as the accurate modeling of the movement of each object and the generation of rainy and snowy environments. Furthermore, the latter challenge requires that physical space in AirSim and WaveFarer and electromagnetic space in Wireless InSite are precisely aligned. For example, the size of scenario, the position of each object, and the weather condition should be  the same in physical space and electromagnetic space.  The developed M$^3$SC dataset  can be accessed from \url{http://pcni.pku.edu.cn/dataset\_1.html}. Also, wireless communication channel data in the M$^3$SC  dataset is  published on the exclusive distributor of Remcom Inc. \url{https://www.renkangtech.com/productinfo/1642251.html?templateId=232054} and \url{https://www.qi-well.com/portal\_c1\_cnt.php?owner\_num=c1\_26875\&button\_num=c1\&folder\_id=22620}.
The major contributions and novelties of this paper are outlined as follows.
\begin{enumerate}
	\item For multi-modal sensing-communication integration, i.e., SoM, a  MMM dataset, named  M$^3$SC, is developed and the corresponding generation framework is proposed \emph{for the first time}. The  M$^3$SC  dataset  aligns physical space and electromagnetic space, and consists of high-fidelity multi-modal  sensory data  and precise ray-tracing-based wireless channel data. Furthermore, different times of the day, various frequency bands, and various weather conditions are considered.
\item To achieve the in-depth integration of sensing
and communications, in the physical space, set the three-dimensional (3D) coordinate of each vehicle frame by frame to support the continuous movement of multi-vehicles. In  the electromagnetic space,  rain and snow models are generated  to simulate rainy and sunny weather. Also,  a large number of internal files are parsed to generate the dynamic scenarios in batch.
\item To precisely align physical  and electromagnetic space,  detect the size and movement trajectory of each object so that they are the same in physical and electromagnetic space. Furthermore, the flexible setting of weather parameters, employment of LISA algorithm \cite{LISA}, and dynamic adjustment of rain and snow models lead to the consistent  weather condition in physical  and electromagnetic space.
\item  At present, the M$^3$SC dataset includes 1500 snapshots, containing 80 RGB images with $1920 \times 1080$ resolution, 160 depth maps with $1920 \times 1080$ resolution, 80 LiDAR point
clouds, 256 sets of mmWave waveforms with 8
radar point clouds, and 72 channel impulse response (CIR) matrices per snapshot, totaling 120,000 RGB images,
240,000 depth maps, 120,000 LiDAR point clouds, 384,000 sets of mmWave waveforms
with 12,000 radar point clouds, and 108,000 CIR matrices. The processing result of the M$^3$SC dataset reveals  the effects of weather, time of the day, and communication frequency, on  sensory and  communication data.
\end{enumerate}

The remainder of this paper is organized as follows. Section II introduces a typical example in the M$^3$SC dataset under vehicular urban crossroads to present the generation framework of the M$^3$SC dataset. Section III illustrates the processing results of sensory  and communication data and gives the valuable  analysis. Section IV discusses the potential MMM sensing-communication application that can be supported by the  M$^3$SC dataset. Finally, Section V draws the conclusion and future work.

\section{Generation framework of the M$^3$SC Dataset: An Example}

\begin{figure*}[!t]
		\centering	\includegraphics[width=0.99\textwidth]{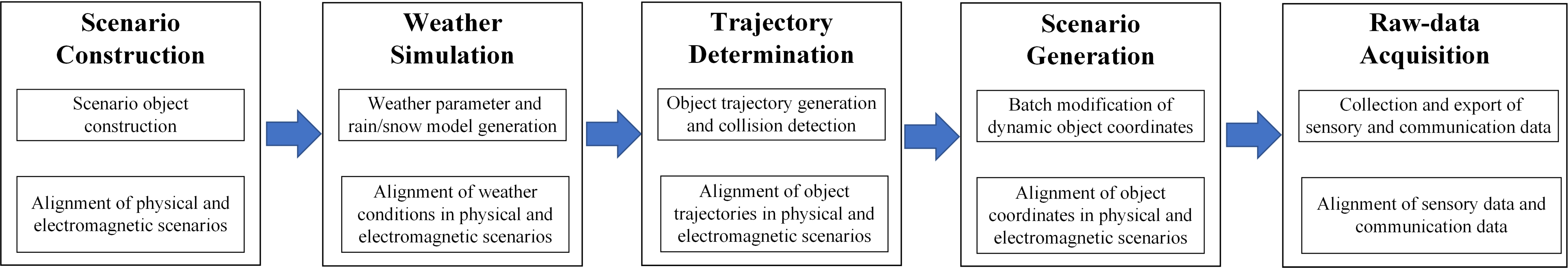}
	\caption{Generation framework of the M$^3$SC dataset.}
	\label{S2_framework}
	\end{figure*}

In this section, a typical example/scenario, i.e., complex high-mobility vehicular urban crossroad scenario, 
in the M$^3$SC dataset is introduced in detail.  Fig.~\ref{S2_framework} gives the  generation framework of the M$^3$SC dataset, which contains five steps and will be discussed in the sequel.
 
\begin{figure*}[!t]
		\centering	\includegraphics[width=0.99\textwidth]{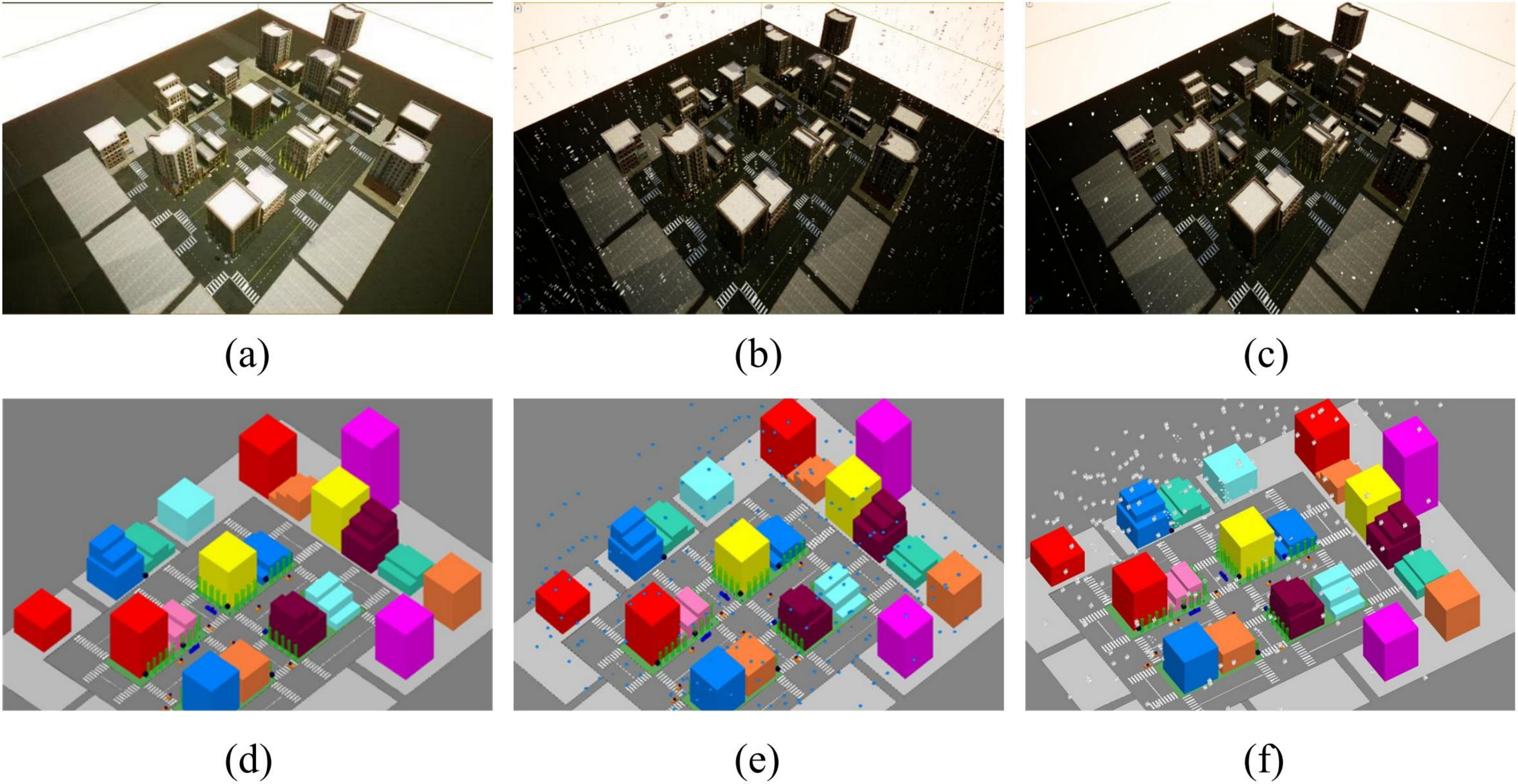}
	\caption{A vehicular urban crossroad environment.  (a) Sunny day in AirSim. (b) Rainy day in AirSim. (c) Snowy day in AirSim. (d) Sunny day in Wireless InSite. (e) Rainy day in Wireless InSite. (f) Snowy day in Wireless InSite. }
	\label{scenario_AirSim}
	\end{figure*}

\subsection{Construction of Simulation Scenarios at the Initial Time}
	
To develop the M$^3$SC dataset, the first step is to construct the simulation scenario with aligned physical space and electromagnetic space
at the  initial time. For physical space in AirSim, we utilize a vehicular urban crossroad scenario produced by PurePolygons, named  Modular Building Set. 
The scenario resolution  is $2048\times 2048$,
which can enable the generation of high-fidelity sensory data.  
Subsequently, the AirSim simulation platform, a plug-in that is constructed on the efficient 3D game engine Unreal Engine, is leveraged to add vehicles and roadside devices into the  vehicular urban crossroad scenario and collect the multi-modal sensory data. The vehicular urban crossroad scenario contains 11 base stations (BSs), 12 vehicles (9 cars and 3 buses), 6 pedestrians, many trees, and various buildings. The BS and vehicle are equipped with multi-modal sensors, including
a camera,  two types of depth cameras, and a LiDAR device.   Furthermore, it is noteworthy that
the parameters of sensing equipment can be properly adjusted, including the field-of-view (FoV), resolution, the range of LiDAR, as well as the number of LiDAR channels, etc. This exceedingly increases the scalability of the sensory data. 

For physical space in WaveFarer, we utilize 3D models  aligned with AirSim, including buildings, trees, pedestrians, and vehicles, ensuring the consistency between the environment features captured by the radars in WaveFarer and the sensors in AirSim. In the vehicular urban crossroad scenario, we introduce ten mmWave radars, including three  roadside radars, three front-view onboard radars, and one vehicle with front-, back-, left-, and right-view onboard radars. Each radar is equipped with 8 transmitter (Tx) antennas and 4 receiver (Rx) antennas, totaling 32 virtual sub-channels.  The radar utilizes a linear frequency modulated continuous wave (FMCW) operating in the frequency range of $77$ GHz to $81$ GHz. The azimuth information is obtained with the help of digital beamforming (DBF), allowing for the acquisition of radar point clouds that encompass azimuth, distance, and Doppler velocity.  The maximum detection range for a certain radar device is $74.9$ m with a range resolution of $0.1499$ m. The Doppler velocity range is $\pm47.42$ m/s with a velocity resolution of $0.939$ m/s. Depending on the requirements of different research tasks, we can flexibly adjust the aforementioned parameters and employ user-defined methods to process radar data to obtain environment features.

For electromagnetic space, we construct a corresponding vehicular urban crossroad communication environment, which precisely aligns with  physical space. Specifically, by omitting
the  fine physical detail, the vehicular urban crossroad scenario constructed in AirSim simulation platform can be imported into 
Wireless InSite simulation platform. The color of each side of the building is modified and the crosswalk is added.
Then,  the accurate wireless communication channel
data can be obtained via the ray-tracing technology.
The BS and vehicle are equipped with a Tx and a Rx. Note that  vehicle-to-vehicle (V2V) links and vehicle-to-infrastructure (V2I) links are considered.
The numbers of antenna elements equipped on each BS and each vehicle are 128 and 32, respectively. The frequency bands cover sub-6 GHz, i.e., carrier frequency is $5.9$ GHz  with $20$ MHz communication bandwidth, and mmWave, i.e., carrier frequency is $28$ GHz  with $2$ GHz communication bandwidth. As mentioned in \cite{huang1,hece}, it is significantly difficult to carry out massive MIMO mmWave channel measurement campaigns under complex high-mobility vehicular communication scenarios. Therefore, the communication data in the  M$^3$SC dataset can fill this gap.
 By adjusting the channel-related parameters, more diverse propagation environments with different antenna elements of transceivers and frequencies can be constructed, leading to  high scalability of the communication data.

The physical space in AirSim and WaveFarer and electromagnetic space in Wireless InSite are precisely aligned after being constructed.
 To be specific, the dimension and coordinate of each object are  compared and aligned in physical and electromagnetic space at the initial time instant to guarantee the aligned sensory  and communication data.

\subsection{Weather Simulation} 

The second step of constructing the M$^3$SC dataset is to carry out the weather simulation.
According to \cite{weather-2,weather-communication}, weather
has a significant effect on  sensory data and communication data. The M$^3$SC dataset includes three typical weather conditions, i.e., sunny, rainy, and snowy days. The weather condition simulation in WaveFarer is currently underway and will be completed in the near future.
Fig.~\ref{scenario_AirSim} shows the scenarios on sunny, rainy, and snowy days in AirSim and Wireless InSite.
For physical space,  we use
Python  to set the parameters of the amount of rain and snow in AirSim. Specifically,  the  parameters of mimicking rainy and snowy days include Rain, Roadwetness, Snow, and RoadSnow, which represent the amount of rain, the degree of water accumulation on the ground, the amount of snow, and the degree of snow accumulation on the ground, respectively. The value of the aforementioned parameters satisfies $[0,1]$. In the simulation, on the rainy day, the Rain and Roadwetness parameters are set to $0.9$ and $1$. Different from the rainy day, Snow and RoadSnow parameters are set to $0.5$ and $1$ on the snowy day.
However, due to the lack of accurate modeling of object properties in AirSim, such as material reflectivity, weather conditions only affect camera sensors while not LiDAR sensors. Such a limitation can also be found in other simulation platforms, such as CARLA \cite{CARLA}, LGSVL \cite{LGSVL}, and DeepGTAV \cite{DeepGTAV}. To overcome this limitation of AirSim, we employ a LiDAR point weather augmentation algorithm, named  LISA \cite{LISA},  to augment points in existing sunny scenarios. The augmentation algorithm proposes a physics-based hybrid Monte Carlo based LiDAR scatterer simulator for adverse weather conditions, such as rainy and snowy days. 
 To leverage the LISA algorithm, we properly set the reflectivity of objects and LiDAR attributes. Furthermore, the parameters controlling rain and snow in the LISA algorithm are rain rate and snow rate, which are set to $50$mm/hr and $10$mm/hr in the simulation according to \cite{LISA}, respectively.
 As a consequence, LiDAR point clouds can include the effects of rain and snow  in a low-complexity manner. Furthermore, the augmented data has been proven to be effective when being utilized to train neural networks, thus ensuring the accuracy of introducing weather effects in LiDAR point clouds. Specifically, to validate the generated LiDAR point clouds on rainy and snowy days, the authors in \cite{LISA} utilized the generated LiDAR point clouds on rainy and snowy days as the training set and exploited three state-of-the-art 3D object detection neural networks to train them. Furthermore, the authors tested  the LiDAR point clouds collected from  rainy and snowy days in reality, i.e.,  Waymo Open Dataset \cite{sensory-1}. The testing result demonstrated that the mean average precision  of the LISA algorithm outperforms that of typical data augmentation algorithms. Therefore, the generated LiDAR point clouds on rainy and snowy days can be  effectively utilized to train neural networks.
 
For the communication data,  rainy  and snowy days in the electromagnetic space can be imitated by introducing rainy and snowy models in Wireless InSite. To be specific, the shapes of rainy and snowy models are set as raindrops and snowflakes, respectively. In addition, on rainy days, the parameters of temperature and humidity  are  set as $22.2$ degrees Celsius and $100$\% in the simulation. Different from rainy days,
the parameters of temperature and humidity  are  set as $-10$ degrees Celsius and $20$\% on snowy days.

The weather conditions of physical space  and electromagnetic space need to be consistent. Towards this objective, we properly set  the weather parameters in  AirSim to align with the weather condition in electromagnetic space. Meanwhile, we accurately modify the rainy and snowy models in electromagnetic space  to align with  physical space before importing them into Wireless InSite. Meanwhile,
the rain and snow parameters in AirSim and the rain and snow rate parameters in the LISA algorithm are adjusted to be consistent. As a result, the M$^3$SC dataset aligns the 
multi-modal sensory data with the wireless communication  data under different  weather conditions.

\begin{figure}[!t]
		\centering	\includegraphics[width=0.49\textwidth]{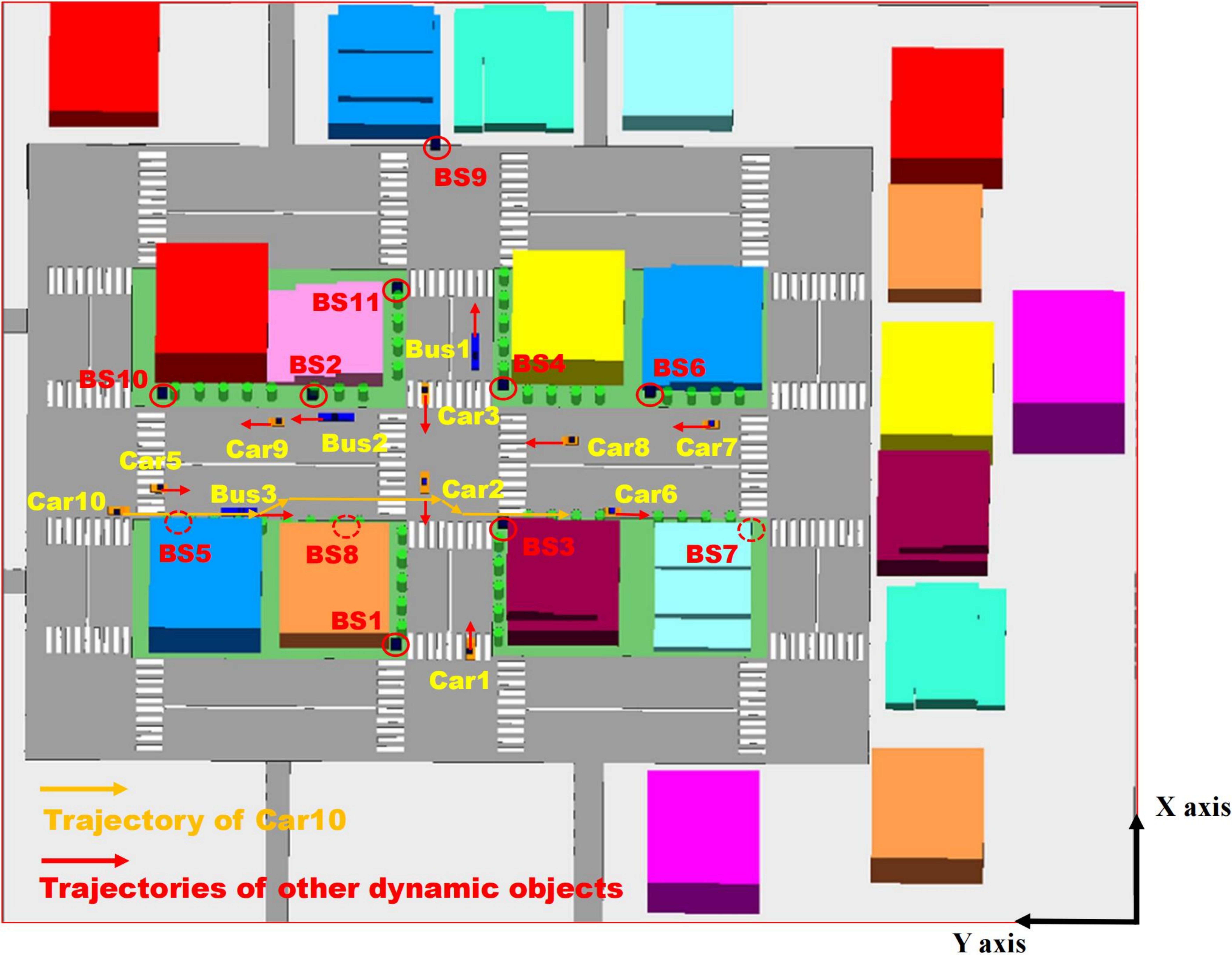}
	\caption{The scenario and  detailed trajectories of  dynamic objects in Wireless InSite simulation platform.}
	\label{trajectory_WI}
	\end{figure}
	
	\begin{table*}[!t]
		\centering
		\caption{The detailed parameter setting of trajectories of dynamic objects.}
		\begin{tabular}{cccccc}
			\hline
			\makecell[c]{\textbf{Object  }\\\textbf{name}}	 &
	\makecell[c]{\textbf{X-axis velocity  }\\\textbf{(m/snapshot)}}	 & 	\makecell[c]{\textbf{Y-axis velocity  }\\\textbf{(m/snapshot)}} &	\makecell[c]{\textbf{Z-axis velocity  }\\\textbf{(m/snapshot)}}& 	\makecell[c]{\textbf{Movement }\\\textbf{start snapshot}}&
		\makecell[c]{\textbf{Movement }\\\textbf{end snapshot}}\\
			\hline
		Car1	& 0.0833 &0 & 0 & 1 & 300 \\ 
			\hline
		Car2	& -0.1 &0 & 0 & 1 & 830 \\
			\hline
	Car3& -0.1 &0 & 0 & 1 & 950 \\
			\hline
	Car5	& 0 &-0.15 & 0 & 1 & 835 \\
			\hline
Car6	& 0 &-0.1 & 0 & 1 & 480 \\
\hline
Car7	& 0 &0.07 & 0 & 1 & 1500 \\
			\hline
Car8	& 0 &0.06 & 0 & 1 & 1500 \\
			\hline
	Car9	& 0 &0.04 & 0 & 1 & 1000 \\
	\hline
\multirow{6}{*}{Car10} 		&	0& -0.1 & 0 & 1& 300 \\	
			\cline{2-6}  
			&	0.1& -0.1 & 0 & 301& 350	\\
			\cline{2-6} 
		&	0& -0.1 & 0 & 351& 660	\\
			\cline{2-6} 
	&	-0.1& 0.1 & 0 & 661& 710	\\
			\cline{2-6}  
		&	0& -0.1 & 0 & 711& 1300	\\
			\cline{2-6}
		&	0& -0.05 & 0 & 1301& 1500	\\
			\cline{2-6}  
			\hline
	Bus1	& 0 &0.05 & 0 & 1 & 320 \\
			\hline
Bus2	& 0 &0.03 & 0 & 1 & 1500 \\
			\hline
	Bus3	& 0 &-0.05 & 0 & 1 & 1500 \\
			\hline
		\end{tabular}	
		\label{trajectory_Setting}
	\end{table*}
 
\subsection{Trajectory Determination  and Collision Detection}  
The third step is to  determine the trajectories of dynamic objects and carry out collision detection based on the determined trajectory. For the precise alignment of physical space and electromagnetic space, the dynamic object has the same trajectory in  AirSim, WaveFarer, and Wireless InSite simulation platforms.
To achieve a decent trade-off between time cost and data volume, we set the number of snapshots in the simulation to 1500.
Due to the same trajectory in the aforementioned three simulation platforms, Fig.~\ref{trajectory_WI} only gives trajectories of dynamic objects in Wireless InSite simulation platform for clarity. Furthermore, a detailed parameter setting of trajectories  is listed in Table~\ref{trajectory_Setting}. 
According to Fig.~\ref{trajectory_WI} and Table~\ref{trajectory_Setting},  the M$^3$SC dataset can support the multi-trajectory and multi-velocity vehicle simulation, which is necessary to be mimicked due to its great impact on sensory and communication data \cite{sensory-2, Huang1}.
 By setting different snapshot values, the M$^3$SC dataset can further contain aligned sensory and communication data under different vehicular velocities.

Based on the  trajectory of the dynamic object, a new collision detection mechanism is introduced to ensure  the validity of the generated complex high-mobility vehicular scenario. If a certain dynamic object collides, its name and the corresponding collision snapshot can be obtained. As a result, the trajectory of the colliding dynamic object can be properly reset to avoid the unrealistic phenomenon.

\subsection{Batch Generation of Dynamic Simulation Scenarios}
\begin{figure*}[!t]
		\centering	\includegraphics[width=0.99\textwidth]{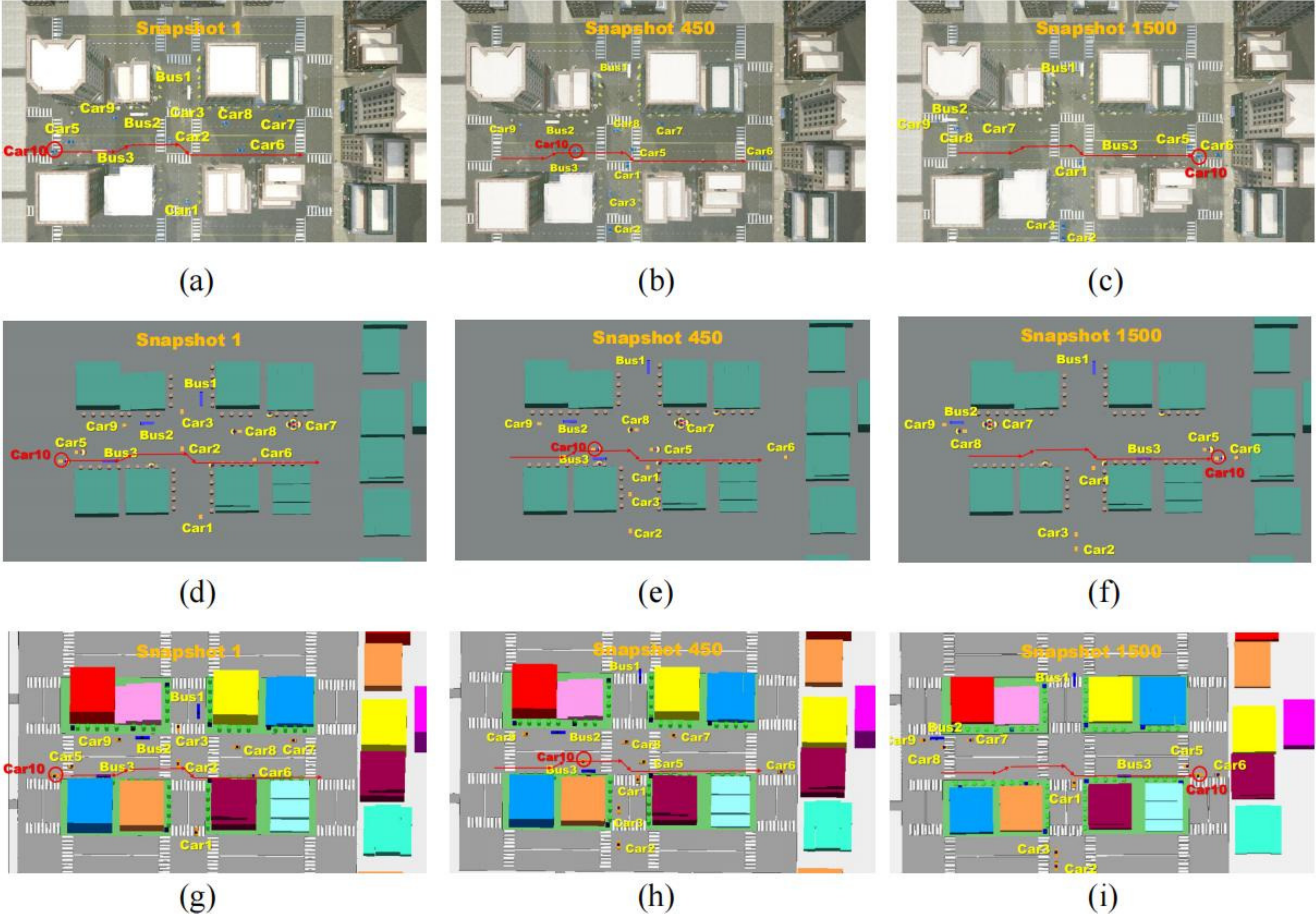}
	\caption{Generated high-mobility vehicular urban crossroad scenarios. (a)-(c) Scenarios in AirSim at Snapshot 1, Snapshot 450, and Snapshot 1500, respectively. (d)-(f) Scenarios in WaveFarer at Snapshot 1, Snapshot 450, and Snapshot 1500, respectively. (g)-(i) Scenarios in Wireless InSite at Snapshot 1, Snapshot 450, and Snapshot 1500, respectively. }
	\label{Scenario_1_1500}
	\end{figure*}
	
Based on the  trajectory of the dynamic object, the fourth step is to complete the batch generation of the vehicular urban crossroad scenario with 1500 snapshots via  efficient simulation software. For physical space in AirSim, positions of dynamic vehicles and sensors are revised in batches via Python. The 3D coordinates of each object and sensor are set frame by frame.
Consequently,
multi-modal sensory data of dynamic vehicles at each snapshot can be collected.  For physical space in WaveFarer, the vehicle and radar positions are controlled via  QtScript  supported by WaveFarer. Based on QtScript scripts, we can import the vehicle positions and orientations  from an external source. Additionally, the relative positions of the onboard radars are fixed to the corresponding vehicles through the coordinate parameter linkage. As a result, radar data related to dynamic vehicles at each snapshot can be collected in WaveFarer.
For the electromagnetic  space, a large number of files that store 3D coordinates of
dynamic vehicles and their antenna elements in Wireless InSite simulation platform are modified in batches via MATLAB. As a result, 1500  vehicular urban crossroad  communication scenarios with different positions of dynamic vehicles
can be generated automatically.

Fig.~\ref{Scenario_1_1500} depicts the vehicular urban crossroad scenarios in AirSim, WaveFarer, and  Wireless InSite simulation platforms at  Snapshot 1, Snapshot 450, and Snapshot 1500. From  Fig.~\ref{Scenario_1_1500}, it can be observed that the positions of the dynamic vehicle in physical space and electromagnetic space at each snapshot are the same.  Furthermore, the dynamic vehicle moves from the start snapshot to the end snapshot  exactly according to the trajectory, e.g., Car10 shown in Fig.~\ref{Scenario_1_1500}, which validates the accuracy of the M$^3$SC dataset.
\begin{table*}[!t]
		\centering
		\caption{Exported V2V and V2I links in a vehicular urban crossroad communication scenario.}
		\begin{tabular}{ccc}
			\hline
		\makecell[c]{\textbf{Transmitter set  }}	 &
	\makecell[c]{\textbf{Receiver set}} & 	\makecell[c]{\textbf{Communication link type}}	  	\\
			\hline
		BS3	& Car2 & V2I \\ 
	\hline
		BS3	& Car3 & V2I \\ 
	\hline
	BS3	& Car5 & V2I \\ 
	\hline
	BS3	& Car10 & V2I \\ 
	\hline
	BS5	& Car5 & V2I \\ 
	\hline
	BS5	& Car10 & V2I \\ 
	\hline
	BS6	& Car6 & V2I \\ 
	\hline
	BS6	& Car7 & V2I \\ 
	\hline
	BS7	& Car5 & V2I \\ 
	\hline
	BS7	& Car10 & V2I \\ 
	\hline
	BS8	& Car5 & V2I \\ 
	\hline
		BS8	& Car10 & V2I \\ 
	\hline
	Bus3	& Car10 & V2V \\ 
	\hline
	Car2	& Car3 & V2V \\ 
	\hline
		Car5	& Car7 & V2V \\ 
	\hline
	Car7	& Car8 & V2V \\ 
	\hline
	Car10	& Bus3 & V2V \\ 
	\hline
	Car10	& Car9 & V2V \\ 
	\hline
		\end{tabular}	
		\label{WI_link}
	\end{table*}

\subsection{Raw-Data Acquisition}
Efficiently acquiring the sensory data and communication data is another challenge and the fifth step aims to complete this challenging work. For physical space, each object, which needs to be moved, is arranged according to the preset coordinates/trajectories. Then, based on the determined parameters and positions of sensors, the sensory data can be directly collected and automatically saved in AirSim and WaveFarer simulation platforms.
For the electromagnetic space,  the generated 1500 vehicular urban crossroad communication scenarios in Wireless InSite  simulation platform are queued for simulation by   a  script. When all 1500 scenarios  are simulated, the corresponding CIR matrices can be \emph{exported automatically} through  Python. For each snapshot,
18 typical CIR matrices related to V2V and V2I links, which are listed in Table~\ref{WI_link}, are successfully exported and stored in  ``.mat'' file. It is worth mentioning that the locations of BS in Wireless InSite simulation platform can be found in Fig.~\ref{trajectory_WI}.
Therefore, in a vehicular urban crossroad communication scenario  at a certain weather condition, e.g., sunny day, with 1500 snapshots,  there are 27,000 CIR matrices, consisting of V2V links together with V2I links. Then, the collected and exported multi-modal sensory data and wireless communication data are aligned.
For clarity, Table~\ref{All_result} properly summarizes the numbers of sensory data, including RGB images, depth maps, LiDAR point clouds, and mmWave radar point clouds, and wireless communication data, i.e., CIR matrices. From Table~\ref{All_result}, it can be readily seen that 
there are currently 120,000 RGB images, 240,000 depth maps,  120,000 LiDAR point clouds, 12,000 mmWave radar point clouds, and 108,000 CIR matrices in the M$^3$SC dataset, covering different weather conditions, different times of the day, as well as different frequency bands. Finally, it is noteworthy that,
by utilizing efficient software whose accuracy is verified by the measurement, the constructed M$^3$SC dataset is precise and similar with the measurement data. Specifically, for sensory data, the urban crossroad scenario in the M$^3$SC dataset is constructed by Unreal Engine, which can provide an excellent rendering effect. In addition, AirSim and WaveFarer can provide high-fidelity multi-modal sensors with adjustable parameters, including FoV degrees, auto exposure speed, number of channels, and number of points captured per second. For communication data, it is obtained based on the ray-tracing technology. Since the ray-tracing technology generates channel data according to the geometrical optics and uniform theory of diffraction, the ray-tracing-based data is precise, which is
similar with the real-world transmission propagation.

\begin{table*}[!t]
		\centering
		\caption{Numbers of RGB images, depth
maps,   LiDAR point clouds, mmWave radar point clouds, and  CIR matrices.}
		\begin{tabular}{cccccc}
			\hline
		\makecell[c]{\textbf{Condition }}	 &
	\makecell[c]{\textbf{RGB images}} & 	\makecell[c]{\textbf{Depth
maps}}	 & \makecell[c]{\textbf{
LiDAR point clouds}} &\makecell[c]{\textbf{
mmWave radar point clouds}} &  \makecell[c]{\textbf{CIR matrices }}	\\
			\hline
	\makecell[c]{Sunny\\Morning/Night\\$5.9$ GHz}	& \multirow{4}{*}{	\makecell[c]{Morning: 30,000\\Night: 30,000}} 		&\multirow{4}{*}{	\makecell[c]{Morning: 60,000\\Night: 60,000}} & \multirow{4}{*}{	\makecell[c]{Morning: 30,000\\Night: 30,000}} & \multirow{4}{*}{	\makecell[c]{Morning: 6,000\\Night: 6,000}}  & 27,000  \\ 
	\cline{1-1} 	\cline{6-6}  
	\makecell[c]{Sunny\\Morning/Night\\$28$ GHz}&	&  &  & & 27,000   \\ 
		\hline
\makecell[c]{Rainy\\Morning\\$28$ GHz}	& 30,000  &  60,000 & 30,000& - & 27,000   \\ 
	\hline	\makecell[c]{Snowy\\Morning\\$28$ GHz}	& 30,000  &  60,000 & 30,000 & - & 27,000   \\ 
	\hline
	\makecell[c]{\textbf{Total}}	& \textbf{120,000}  &  \textbf{240,000}& \textbf{120,000}& \textbf{12,000} & \textbf{108,000}   \\ 
	\hline
		\end{tabular}	
		\label{All_result}
	\end{table*}
	
\section{Results and Analysis}
The  multi-modal sensory data collected in AirSim and WaveFarer simulation platforms and communication data collected in  Wireless InSite simulation platform are presented and analyzed. In addition, the influences of weather, time of the day, and communication frequency on multi-modal sensory and communication data are investigated. Also, the precise alignment
between sensory data and communication data is demonstrated. 

\subsection{Multi-Modal Sensory data}
For the simulation of multi-modal sensory information, four typical sensing cases, i.e., sunny
days in the night, sunny days in the morning, rainy days in the morning, and snowy days in the
morning, are considered.

\begin{figure}[!t]
		\centering	\includegraphics[width=0.499\textwidth]{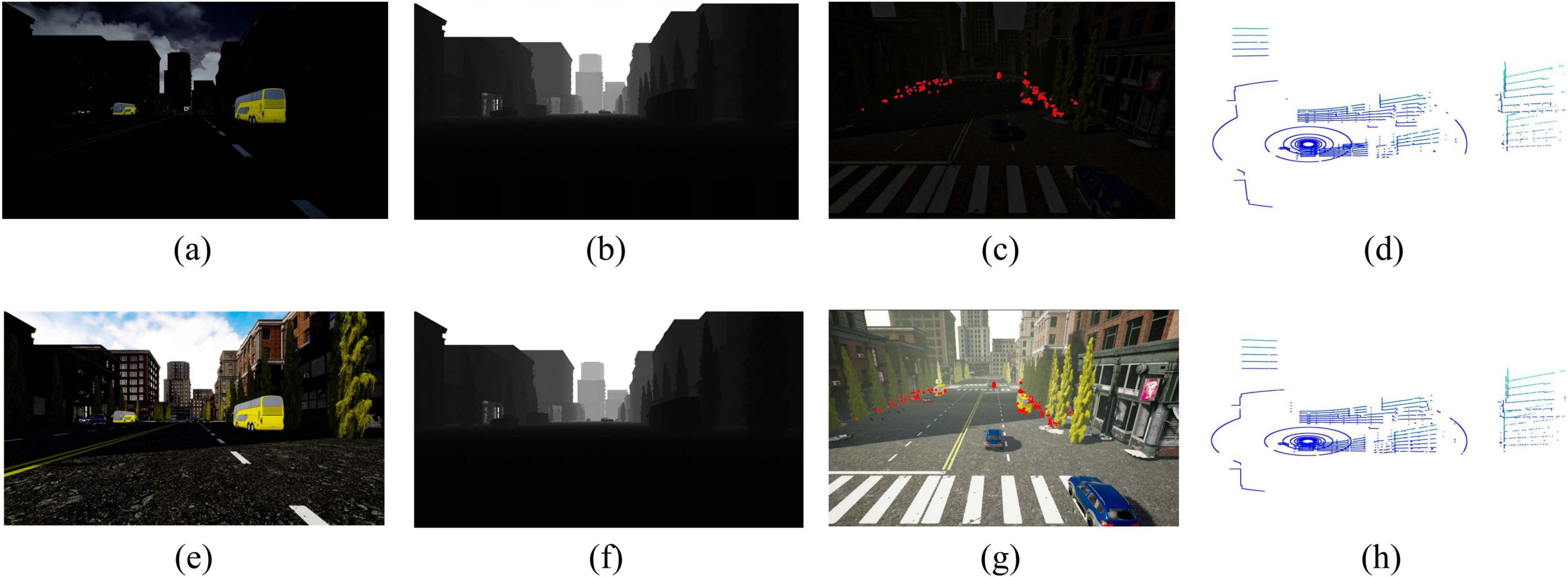}
	\caption{The sensory data collected by Car5 at Snapshot 1 on the sunny day in the morning and night. (a)-(d) RGB image, depth map, mmWave radar point clouds, and   LiDAR point clouds in the night, respectively. (e)-(h) RGB image, depth map, mmWave radar point clouds, and   LiDAR point clouds in the morning, respectively.}
	\label{time_impact}
	\end{figure}
	
Fig.~\ref{time_impact} illustrates the RGB images, depth maps, mmWave radar point clouds, and LiDAR point clouds, on the sunny day at Snapshot 1 in the morning and night. It can be observed that the visibility of RGB image in the night is lower than that in the morning.  This is due to the poor robustness of cameras to sunlight. On the contrary, depth maps and LiDAR point clouds in the night  are similar to those in the morning. This is because that the robustness of depth cameras and LiDAR devices to sunlight is high. Specifically, depth cameras and LiDAR devices utilized in AirSim operate in the near infrared region, and thus are less affected by sunlight and are  insensitive  to day and night. Similar to depth cameras and LiDAR devices, since the robustness of mmWave radar  to sunlight is also high, mmWave radar  point
clouds in the night are similar to those in the morning.

\begin{figure}[!t]
		\centering	\includegraphics[width=0.499\textwidth]{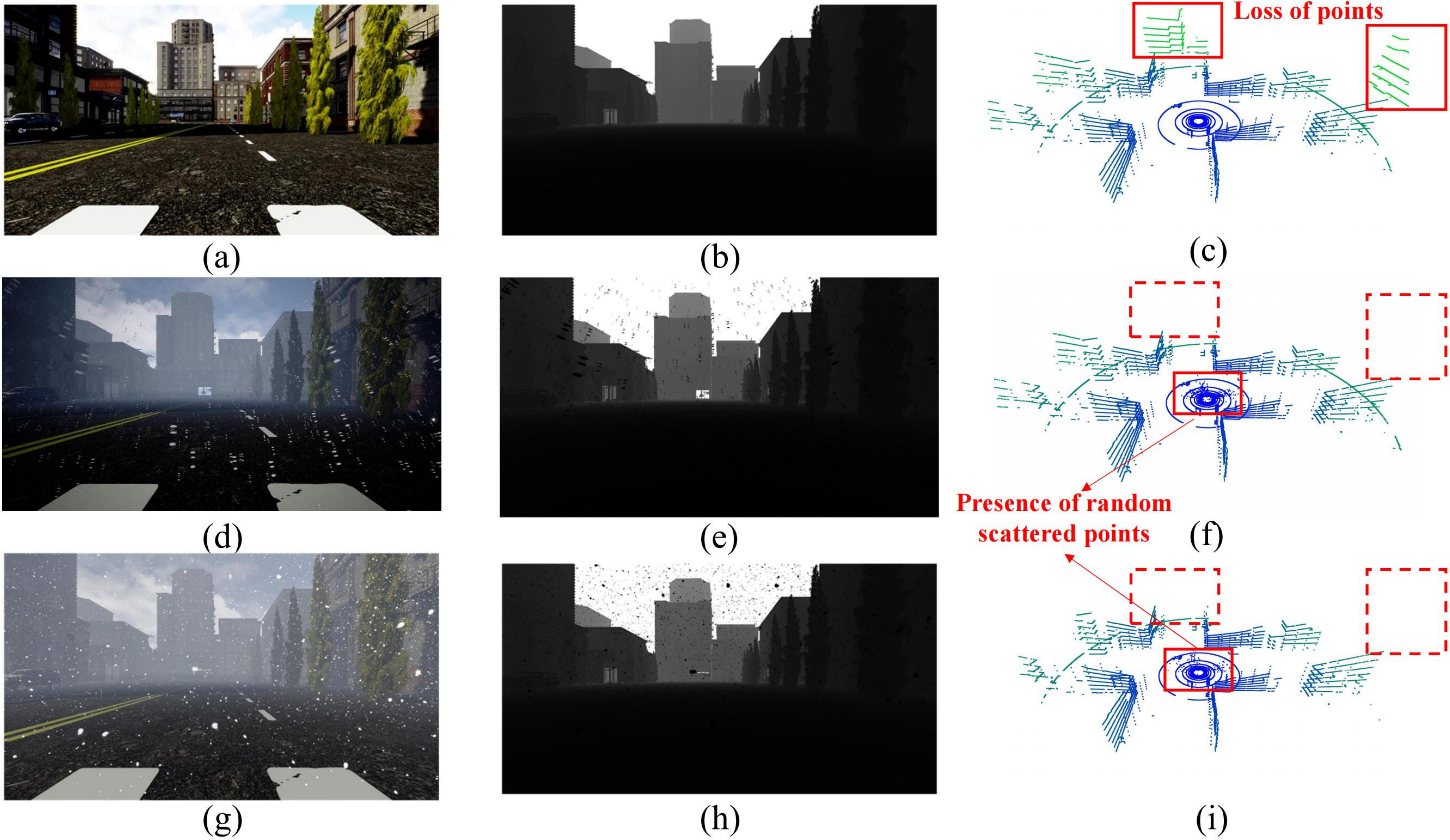}
	\caption{The sensory data collected by Car5 at Snapshot 450 in the morning on the sunny, rainy, and snowy days. (a)-(c) RGB image,  depth map, LiDAR point clouds on the sunny day, respectively. (d)-(f) RGB image,  depth map, LiDAR point clouds on the rainy day, respectively. (g)-(i) RGB image,  depth map, LiDAR point clouds on the snowy day, respectively.}
	\label{weather_impact}
	\end{figure}
	
Fig.~\ref{weather_impact} presents
the RGB images, depth maps, and LiDAR point clouds in the morning at Snapshot 450 on the sunny, rainy, and snowy days. Compared to the sunny day, RGB images and depth maps have lower visibility on the rainy and snowy days, which is consistent with the observation in \cite{weather-2,weather-22}. In addition, LiDAR points are sensitive to rainy and snowy weather conditions. According to laser attenuation caused by random scattering medium, range uncertainty is increased and points far away from LiDAR are lost, resulting in possible vague and missing points within the real target \cite{LISA}. In addition, large back scattered power from random droplets causes the presence of random \emph{scattered points} nearby the LiDAR sensors.

In summary, according to Figs.~\ref{time_impact} and \ref{weather_impact}, it can be concluded that the processing result of multi-modal sensory data is realistic, which verifies the acquired multi-modal sensory data in the M$^3$SC dataset.

\subsection{Wireless Communication Information}
For the simulation of wireless communication information, 
four typical scenarios, i.e., sunny days at sub-6 GHz, sunny days at mmWave, rainy days at mmWave, and  snowy days at mmWave, are taken into account. Important channel statistical properties,  heat maps, and propagation paths under these four typical scenarios are presented.

A basic function, which describes the wireless communication channel, is CIR $h(\varepsilon, t, \tau)$ with
space $\varepsilon$, time $t$, and delay $\tau$. By taking the Fourier transform of CIR $h(\varepsilon, t, \tau)$ with respect of
delay $\tau$, the spatial-temporal varying transfer function $\phi(\varepsilon, t, f)$ is represented by
	\begin{equation}
		\phi(\varepsilon, t, f)=\int h(\varepsilon, t, \tau) e^{-j 2 \pi \tau f} \mathrm{d} \tau
	\end{equation}
where $\int\{\cdot\}$ denotes the integral operator.
Then, the space-time-frequency correlation function (STF-CF)  $\rho_{\phi}(\varepsilon, t, f ; \Delta \varepsilon, \Delta t, \Delta f)$ is expressed by
	\begin{equation}
		\begin{aligned}
		&	\rho_{\phi}(\varepsilon, t, f ; \Delta \varepsilon, \Delta t, \Delta f)\\
  &~~ =\mathbb{E}\left[\phi(\varepsilon, t, f) \phi^{*}(\varepsilon-\Delta \varepsilon, t-\Delta t, f-\Delta f)\right]
			\label{eq:phi}
		\end{aligned}
	\end{equation}
where $\mathbb{E}\{\cdot\}$  represents the expectation operator. Based on the derived STF-CF, channel statistical properties, including space  cross-correlation function (SCCF), time auto-correlation function (TACF), and  Doppler
power spectrum density (DPSD), can be obtained. For the channel, the relationship between the multipath effect and SCCF, TACF, and DPSD, can be given below.
The SCCF reflects the spatial
correlation between different sub-channels with different paths. The TACF describes the temporal correlation of the channel with the appearance and disappearance of paths. The DPSD represents the  received power distribution in the channel with different paths under different Doppler frequency shifts.

\begin{figure}[!t]
		\centering	\includegraphics[width=0.43\textwidth]{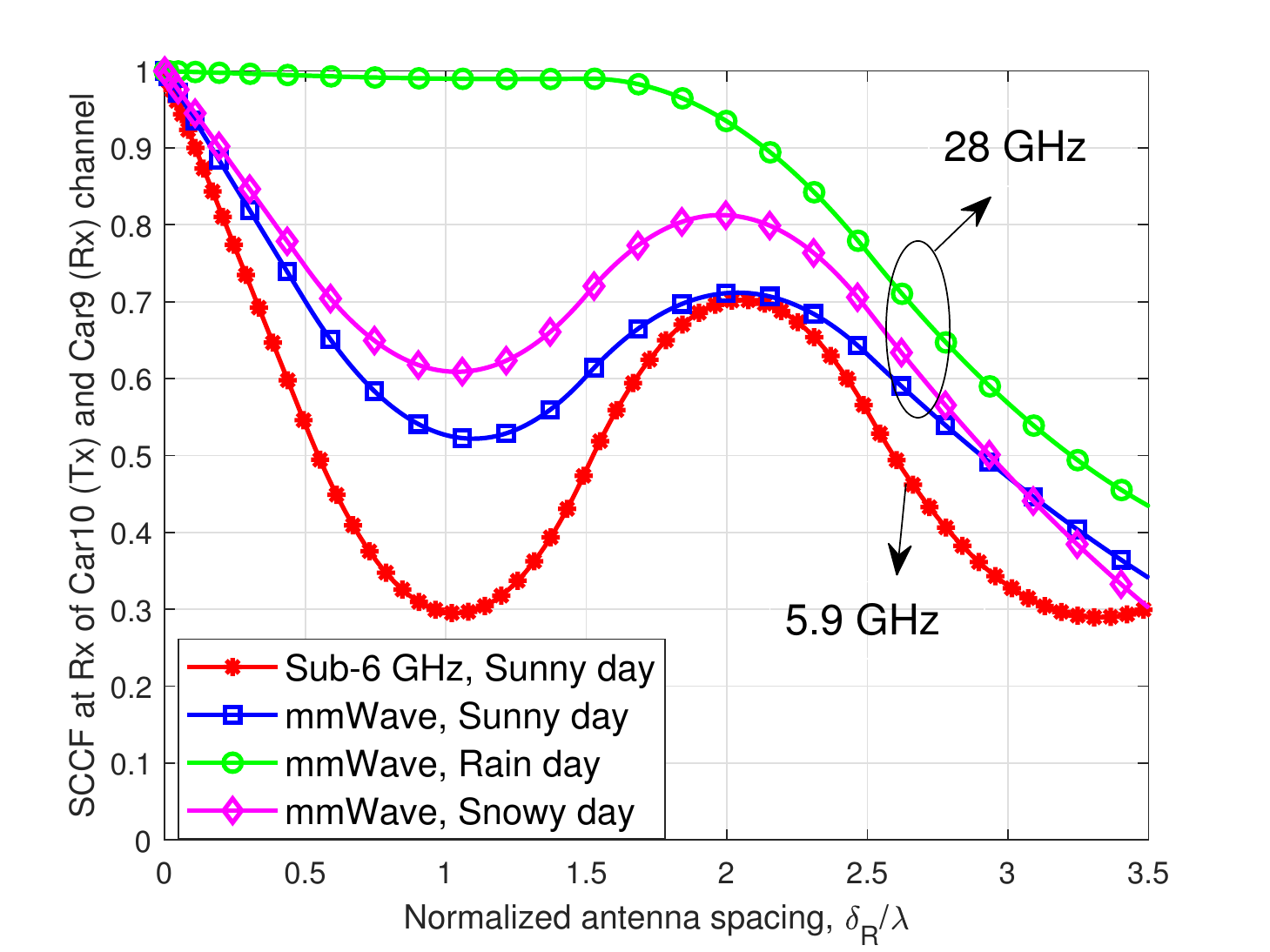}
	\caption{The normalized absolute SCCF of the channel related to Car10 and Car9.}
	\label{SCCF}
	\end{figure}
 By setting $\Delta t=\Delta f=0$, the STF-CF $\rho_{\phi}(\varepsilon, t, f ; \Delta \varepsilon, \Delta t, \Delta f)$ can be simplified to the SCCF. 	
	Fig.~\ref{SCCF} presents the normalized absolute Rx SCCF, where Car10 and Car9, as shown in Fig.~\ref{Scenario_1_1500}(c), are the Tx and the Rx, respectively. Car10 and Car9 are two typical mobile vehicles equipped with large-scale antenna array in the vehicular urban crossroad scenario, and thus the V2V link between them can be viewed as an example in the simulation.
 In 	Fig.~\ref{SCCF}, $\delta_\mathrm{R}$ is  the adjacent antenna spacing at Rx and $\lambda$ is  the carrier wavelength.
Compared to the mmWave frequency band, the channel at sub-6 GHz exhibits a lower SCCF, which is consistent with the measurement result in \cite{Huangj}. This is because that, in comparison of mmWave, 
the multipath effect in  sub-6 GHz is more obvious and channel spatial diversity is larger. Furthermore, the SCCF of mmWave channels on the sunny day is lower than that on the rainy and snowy days.  The underlying physical reason is that, on  rainy and snowy days, the propagation power attenuation is more significant, resulting in less obvious  multipath effect and larger spatial correlation.

\begin{figure}[!t]
		\centering	\includegraphics[width=0.43\textwidth]{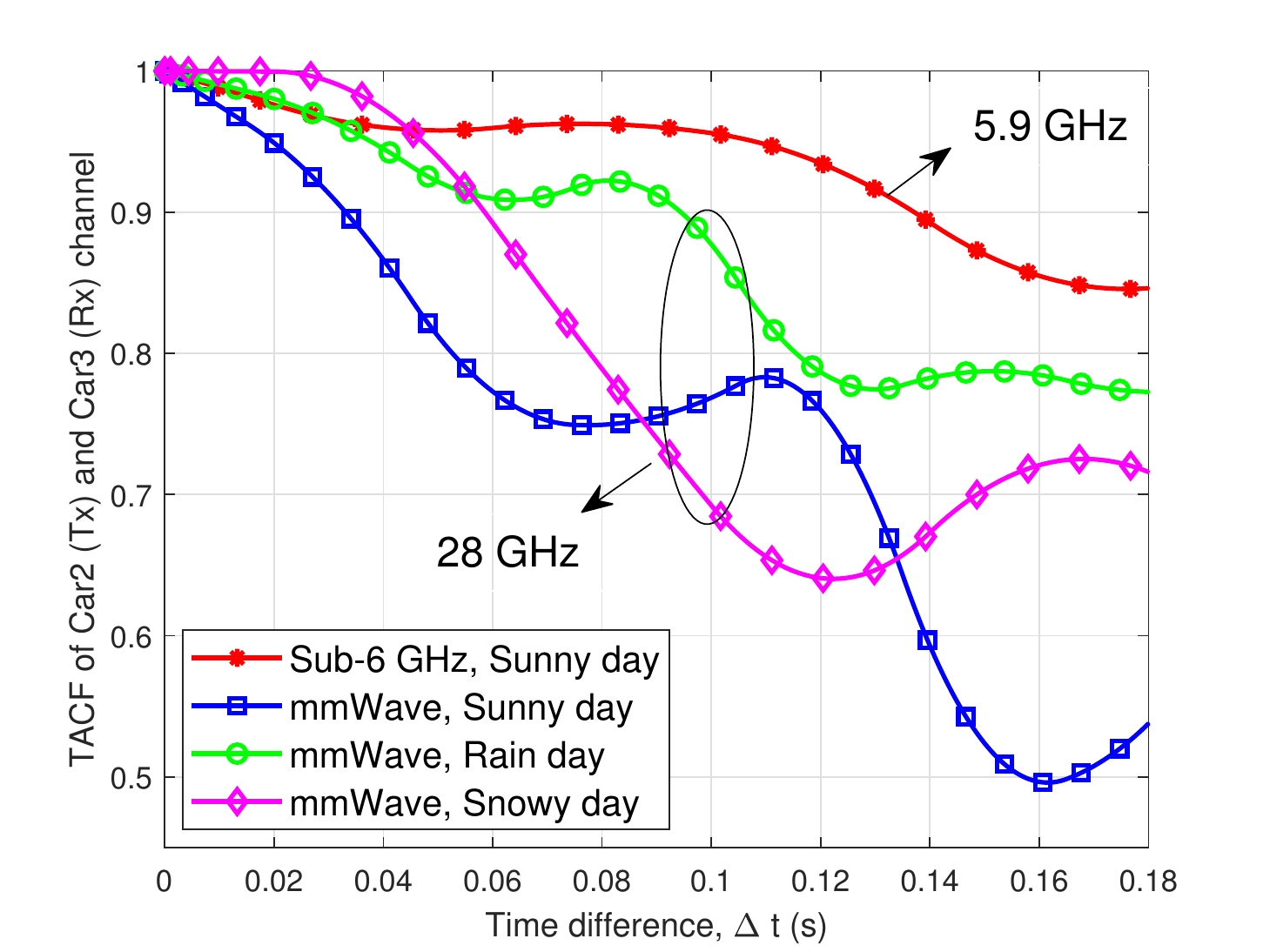}
	\caption{The normalized absolute TACF of the channel related to Car2 and Car3.}
	\label{TACF}
	\end{figure}
By setting $\Delta \varepsilon=\Delta f=0$ in the STF-CF $\rho_{\phi}(\varepsilon, t, f ; \Delta \varepsilon, \Delta t, \Delta f)$, the TACF is obtained.
Fig.~\ref{TACF} depicts the normalized absolute TACF, where Car2 and Car3, as shown in Fig.~\ref{Scenario_1_1500}(c), are the Tx and the Rx, respectively. The TACF at mmWave frequency band is significantly lower than that at sub-6 GHz frequency band. The philosophy is that, compared to sub-6 GHz,
Doppler spread at mmWave is  severer and the channel coherence time is much smaller \cite{stuber}. Furthermore, the mmWave channel on the sunny day exhibits a lower TACF than that on the rainy and snowy days. This is because that there are more multipaths on the  sunny day and the corresponding channel is more rapidly-changing. In this case, the appearance and disappearance of paths are rapider and more complex over time. As a consequence, the temporal correlation of the channel at different time instants is smaller, resulting in a lower TACF.

\begin{figure}[!t]
		\centering	\includegraphics[width=0.43\textwidth]{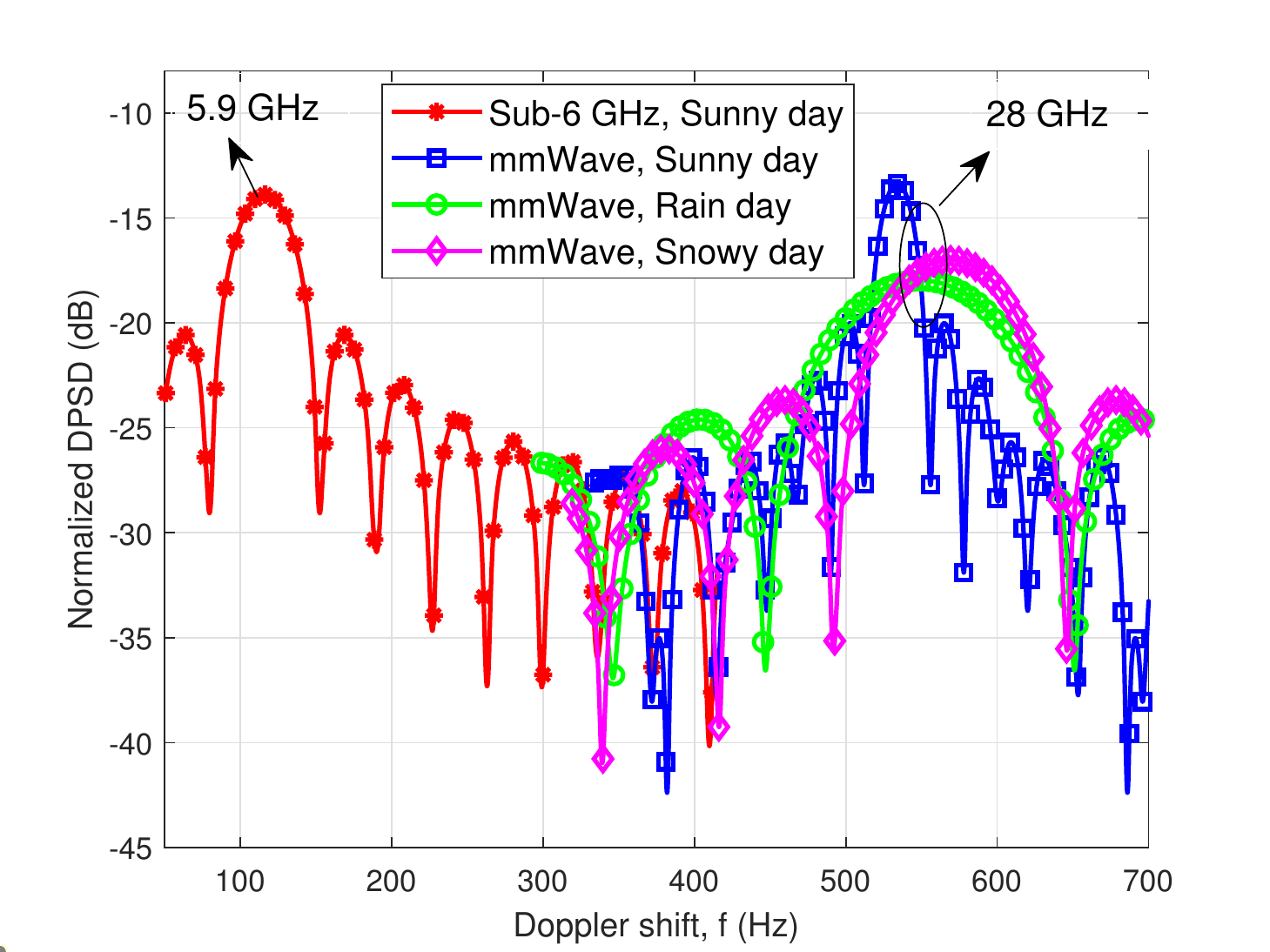}
	\caption{The normalized DPSD of the channel related to Car10 and Car9.}
	\label{DPSD}
	\end{figure}
Through the Fourier transfer of the TACF with respect to time interval $\Delta t$, the Doppler PSD is obtained.
In Fig.~\ref{DPSD}, the normalized DPSD related to the Car10 and Car9 channels is shown. Attributed to the higher  frequency in  mmWave communications, the channel at mmWave  has a larger Doppler spread than that at sub-6 GHz \cite{stuber,TCOMmy}. Additionally, in comparison with rainy and snowy days, the channel on the sunny day has a steeper distribution of DPSD. This is due to the fact that the line-of-sight (LoS) component on the sunny day is more dominant than that on the rainy and snowy days. 

\begin{figure}[!t]
		\centering	\includegraphics[width=0.43\textwidth]{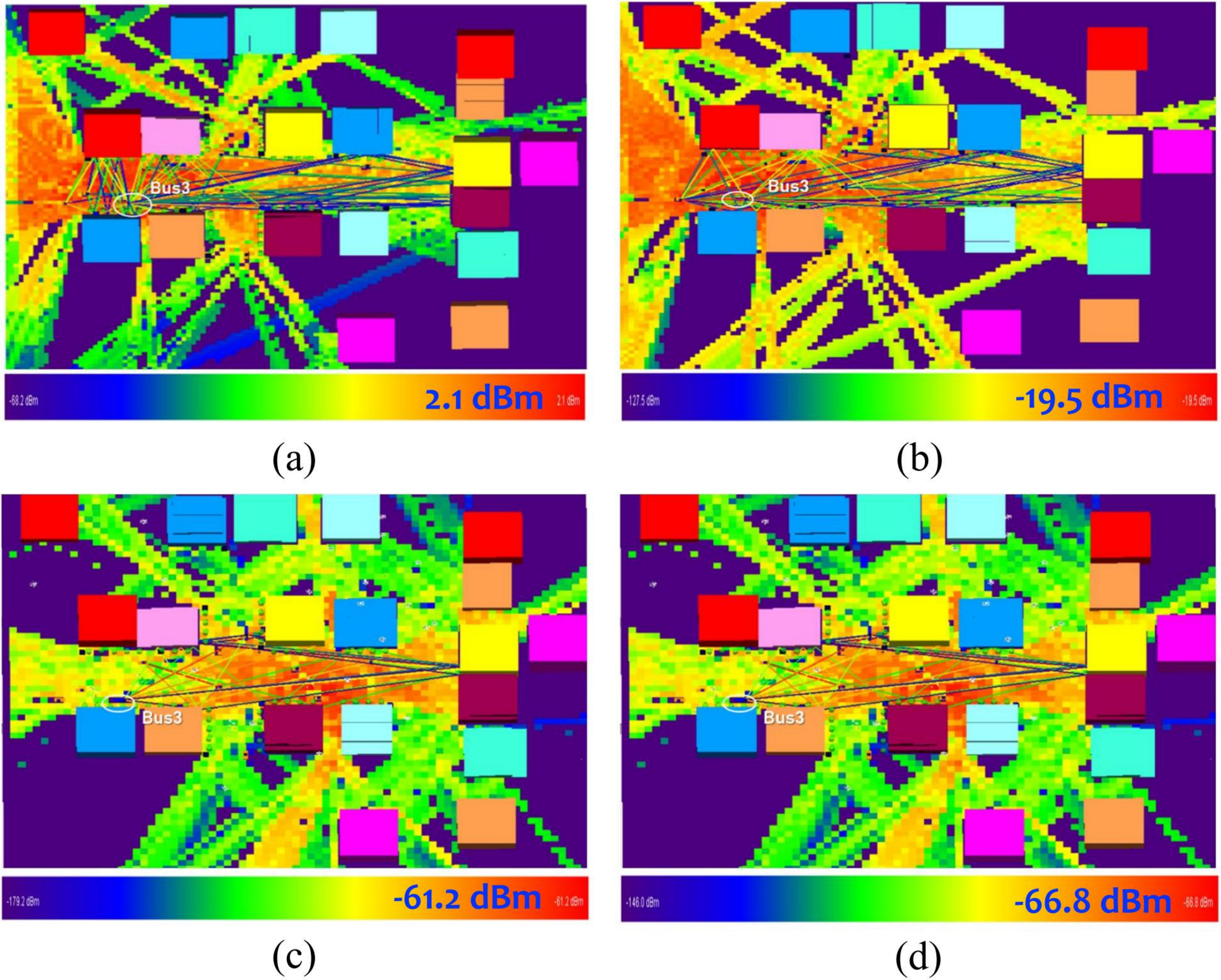}
	\caption{Heat maps and propagation paths in Wireless InSite simulation platform. (a) Sunny day at sub-6 GHz. (b) Sunny day at mmWave. (c) Rainy day at mmWave. (d) Snowy day at mmWave.}
	\label{Heatmap}
	\end{figure}
	
Fig.~\ref{Heatmap} gives the heat map and the propagation path related to Bus3. 
To obtain the heat map, by utilizing the visualization function in Wireless InSite, the received power via the planar antenna array, i.e., $xy$-grid, is displayed in the form of the heat map. For the propagation path, Wireless InSite can also show the ray-tracing-based propagation paths  in the environment via the visualization function.
By obtaining the heat map and the propagation path under different weather conditions and frequency bands, some valuable observations can be found. Compared to the sub-6 GHz frequency band, the path power and the number of propagation paths are much smaller at the mmWave frequency band due to the significantly high pathloss. In addition, compared to the sunny day, the path power and the number of propagation paths are much smaller on the rainy and snowy days owing to the  obvious rain and snow attenuation at mmWave communications \cite{comst-weather}.

According to Figs.~\ref{SCCF}-\ref{Heatmap}, it can be concluded that the processing result of wireless communication data is accurate, which validates the wireless communication data in the M$^3$SC dataset.

\subsection{Aligned Sensory Data and Communication Data}
\begin{figure}[!t]
		\centering	\includegraphics[width=0.43\textwidth]{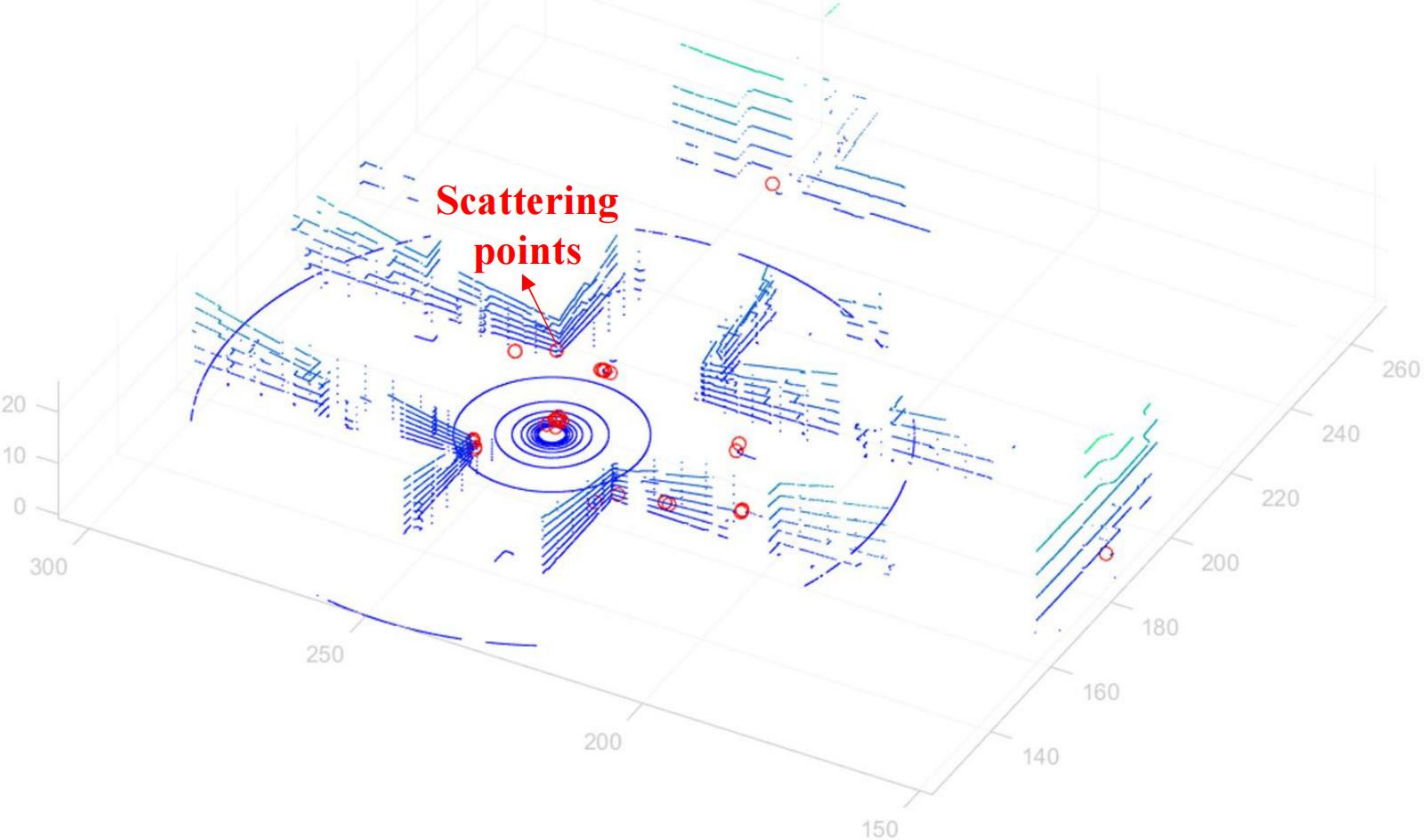}
	\caption{Mapping between LiDAR point clouds and scattering points at the initial time.}
	\label{match}
	\end{figure}
Fig.~\ref{match} shows  LiDAR point clouds in physical space and  scattering points in electromagnetic space  on the sunny day at the initial time. For LiDAR point clouds,  Car1 equips with the LiDAR device. For the scattering point,  Car1 and Car2 are the Tx and Rx, respectively, where the carrier frequency is $28$ GHz  with $2$ GHz communication bandwidth.   From Fig.~\ref{match}, it can be readily observed that the M$^3$SC dataset has the aligned LiDAR point clouds  and scattering points. Specifically, the scattering point exists on the vehicle, building, and trees detected by  LiDAR point clouds. This is consistent with the realistic propagation environment.

Based on Fig.~\ref{match}, it can be concluded that the aligned sensory data and communication data are obtained. Therefore, physical space and electromagnetic
space are precisely aligned in
the  M$^3$SC dataset. 

\section{Potential MMM Sensing-Communication Applications Supported by M$^3$SC Dataset}
In this section, the potential MMM sensing-communication applications that can be supported by the comprehensive M$^3$SC dataset are  discussed  from three perspectives.

\begin{enumerate}
\item \emph{Channel modeling:} The M$^3$SC dataset can be regarded as an important component for the development of novel channel modeling methods. With the help of aligned multi-modal sensory  and communication data in the M$^3$SC dataset,  the correspondence between objects in physical space and clusters in electromagnetic space can be explored. Meanwhile, the complicated correlation between object parameters and cluster parameters can also be obtained via  deep learning algorithms. Furthermore, the obtained correlation can be extended to various weather conditions, times of the day, and frequency bands via the M$^3$SC dataset.
\item \emph{Waveform design:} The M$^3$SC dataset can support new paradigms of waveform design for more in-depth integration of sensing and communications by providing multi-modal sensing information and accurate real-time channel information. Specifically, the correlation between physical space and electromagnetic space can be utilized to achieve mutual benefits of sensing and communications. The mapping from the information in physical space to the optimal integrated waveform can be explored. The optimal integrated waveform can be used to further enhance the sensing accuracy and robustness of conventional sensors. Additionally, a more effective extraction of electromagnetic space can increase the secrecy rate for security communications. Finally, 
by M$^3$SC dataset  with different weather conditions and different frequency bands, the robustness and real efficiency of the waveform design methods can be evaluated adequately.
\item \emph{Perception enhancement:} The synchronized multi-modal and multi-vehicle sensing data in the M$^3$SC dataset can support research on perception system enhancement to address the problem of single-vehicle perception in occlusion or harsh weather \cite{sijiang}.
Based on the M$^3$SC dataset, the information redundancy brought by multiple sensors and vehicles can be quantified. Then, communication resource planning can be carried out to adequately utilize  communications as the method of information transmission. The perception algorithm on different weather and different road conditions can also be trained and tested  the M$^3$SC dataset. The aligned channel state information and sensory data in the M$^3$SC dataset can support the extraction of physical space localization and perception information from the channel state information of V2V and V2I links. Especially in situations where the sensor's information is insufficient, such as camera in dim light and LiDAR on snowy days. 
\end{enumerate}

\section{Conclusions and Future Work}
This paper has developed a    MMM  dataset, named  M$^3$SC, and the generation framework of MMM datasets for  sensing-communication integration. 
The  M$^3$SC dataset has achieved the in-depth integration
of sensing and communications and precise alignment of physical space
and electromagnetic space.
Furthermore, the M$^3$SC dataset has contained three weather conditions, i.e., sunny, rainy, and snowy  days, different times of the day, i.e., morning
and night, and  various frequency bands, i.e.,  sub-6 GHz and mmWave. Currently, there have been 120,000 RGB
images, 240,000 depth maps, 384,000 sets of mmWave waveforms
with 12,000 radar point cloud, 120,000 LiDAR point clouds, and 108,000 CIR
matrices in the M$^3$SC dataset. The data processing result has shown that the weather, time, and frequency bands have significant impacts on  sensory data and communication data. Thanks to the judiciously designed data generation framework and orchestrated cooperation of several simulation platforms, the data processing result has also demonstrated the accuracy of the aligned sensory and communication data.  Finally, this paper has discussed the potential MMM sensing-communication applications with the aid of M$^3$SC dataset from three perspectives.
Currently, we are constructing a realistic measurement platform for sensing and communication integration. In the future, the obtained measurement data can be utilized as a reference for the  M$^3$SC  dataset.
Furthermore, we can  enrich the M$^3$SC dataset by jointly considering different vehicular traffic densities and diversified simulation scenarios, including high-speed train,
unmanned aerial vehicle, maritime scenarios, and warfare scenarios.

\section*{Acknowledgment}
The authors would like to thank Mengyuan Lu and Zengrui Han for their help in the construction of electromagnetic space in Wireless InSite simulation platform and Weibo Wen, Qi Duan, and Yong Yu for their help in the construction of physical space in AirSim simulation platform.

	\ifCLASSOPTIONcaptionsoff
	\newpage
	\fi

	%
	
	%
	%
	%
	
	
	


\begin{thebibliography}{29}
		
		\bibitem{nature}
		S. Dang, O. Amin, B. Shihada, and M.-S. Alouini, ``What should 6G be?'' \emph{Nat. Electron.}, pp. 20--29, Jan. 2020. 

\bibitem{comst-1}
X. Cheng, Z. Huang, and L. Bai, ``Channel nonstationarity and consistency for beyond 5G and 6G: A survey,'' \emph{IEEE Commun. Surveys Tutor.}, vol. 24, no. 3, pp. 1634--1669, third-quarter 2022.

\bibitem{comst-2}
D. Serghiou, M. Khalily, T. W. C. Brown, and R. Tafazolli, ``Terahertz channel propagation phenomena, measurement techniques and modeling for 6G wireless communication applications: A survey, open challenges and future research directions," \emph{IEEE Commun. Surveys Tutor.}, to be published, 2022. Doi: 10.1109/COMST.2022.3205505.

\bibitem{TWCMY}
X. Cheng, Z. Huang, and S. Chen, ``Vehicular communication channel
measurement, modelling, and application for beyond 5G and 6G,'' \emph{IET
Commun.}, vol. 14, no. 19, pp. 3303--3311, Nov. 2020.
		




\bibitem{ISAC-1}
X. Cheng, D. Duan, S. Gao, and L. Yang, ``Integrated sensing and communications (ISAC) for vehicular communication networks (VCN)," \emph{IEEE Internet Things J.}, vol. 9, no. 23, pp. 23441--23451, Dec. 2022.
\bibitem{ISAC-JCIN}
C. Liu, X. Liu, S. Li, W. Yuan, and D. W. K. Ng, ``Deep CLSTM for predictive beamforming in integrated sensing and communication-enabled vehicular networks,'' \emph{J. Commun. Inf. Net.}, vol. 7, no. 3, pp. 269--277, Sep. 2022.

\bibitem{ISAC-IET}
 M. L. Rahman, J. A. Zhang, X. Huang, Y. J. Guo, and Z. Lu, ``Joint communication and radar sensing in 5G mobile network by compressive sensing,'' \emph{IET Commun.}, vol. 14, no. 22, pp. 3977--3988, Feb. 2020.



\bibitem{ISAC-2}
J. Wang \emph{et al.}, ``Integrated sensing and communication: Enabling techniques, applications, tools and datasets, standardization, and future directions," \emph{{IEEE} Internet. Things. J.}, to be published, 2022. Doi: 10.1109/JIOT.2022.3190845.


\bibitem{Data-5}
A. Geiger, P. Lenz, C. Stiller, and R. Urtasun, ``Vision meets robotics: The KITTI dataset,'' \emph{Int. J. Robot. Res.}, vol. 32, no. 11, pp. 1231--1237, Aug. 2013.

\bibitem{WALDO}
Steve Blandino (2021), Dataset of channels and received IEEE 802.11ay signals for sensing applications in the 60GHz band, National Institute of Standards and Technology, https://doi.org/10.18434/mds2-2417 (Accessed 2022-11-07)

\bibitem{Data-7}
 A. Alkhateeb \emph{et al.},
“DeepSense 6G: A large-scale real-world multi-modal sensing and communication dataset,” \emph{available on arXiv}, 2022. [Online]. Available:
https://www.DeepSense6G.net
\bibitem{Data-2}
A. Alkhateeb, ``DeepMIMO: A generic deep learning dataset for millimeter wave and massive MIMO applications,” in \emph{Proc. Inf. Theory Appl. Workshop (ITA)}, San Diego, CA, USA, Feb. 2019, pp. 1--8.

\bibitem{LiDARsim}
 S. Manivasagam \emph{et al.}, ``LiDARsim: Realistic LiDAR simulation by
leveraging the real world,'' in \emph{Proc. IEEE/CVF Conf. Comput. Vis.
Pattern Recognit. (CVPR)}, Seattle, WA, USA, Jun. 2020, pp. 11167--11176.
\bibitem{OPV2V}
R. Xu \emph{et al.}, ``OPV2V:
An open benchmark dataset and fusion pipeline for perception with vehicle-to-vehicle communication,'' \emph{arXiv preprint
arXiv:2109.07644}, 2021.

\bibitem{SHIFT}
T. Sun \emph{et al.}, ``SHIFT: A synthetic driving dataset for continuous multi-task domain adaptation," in \emph{Proc. IEEE/CVF Conf. Comput. Vis.
Pattern Recognit. (CVPR)}, New Orleans, LA, USA, Jun. 2022, pp. 21339--21350.
\bibitem{WI}
\emph{Remcom.} Wireless InSite. [Online]. Available: https://www.remcom.com/wireless-insite-em-propagation-software [Accessed date: Mar. 2022].
\bibitem{Ali}
A. Ali, N. González-Prelcic, and A. Ghosh, ``Passive radar at the roadside unit to configure millimeter wave vehicle-to-infrastructure links," \emph{IEEE Trans. Veh. Technol.}, vol. 69, no. 12, pp. 14903--14917, Dec. 2020.


\bibitem{LIDARS}
A. Klautau \emph{et al.}, ``5G MIMO data for machine learning: Application
to beam-selection using deep learning,'' in \emph{Proc. Inf. Theory Appl.
Workshop (ITA)}, San Diego, CA, USA, Feb. 2018, pp. 1--9.

\bibitem{Klautau}
		A.~Klautau, N.~Gonz{\'a}lez-Prelcic, and R.~W. Heath, ``LIDAR Data for Deep Learning-Based mmWave Beam-Selection,'' {\em IEEE Wireless Commun. Lett.}, vol.~8, no.~3, pp.~909--912, Jun.~2019.



\bibitem{ViWi}
M. Alrabeiah, A. Hredzak, Z. Liu, and A. Alkhateeb, ``ViWi: A deep learning dataset framework for vision-aided wireless communications," in \emph{Proc. IEEE Veh. Technol. Conf. (VTC2020-Spring)}, Antwerp, Belgium, May 2020, pp. 1--5.


\bibitem{AirSim}
 S. Shah, D. Dey, C. Lovett, and A. Kapoor, ``AirSim: High-fidelity
visual and physical simulation for autonomous vehicles,'' in \emph{Field and
Service Robotics}, M. Hutter and R. Siegwart, Eds. Cham, Switzerland:
Springer, 2018, pp. 621--635.


\bibitem{WaveFarer}
\emph{Remcom}, WaveFarer. [Online]. Available: https://www.remcom.com/wavefarer-automotive-radar-software [Accessed date: Mar. 2022].

\bibitem{LISA}
 V. Kilic, D. Hegde, V. Sindagi, A. B. Cooper, M. A. Foster, and
V. M. Patel, ``LiDAR light scattering augmentation (LISA): Physics-based
simulation of adverse weather conditions for 3D object detection,'' 2021, \emph{arXiv:2107.07004}.



\bibitem{huang1}
Z. Huang \emph{et al.}, ``A non-stationary 6G V2V channel model with continuously arbitrary trajectory,'' \emph{IEEE Trans. Veh. Technol.},  vol. 72, no. 1, pp. 4--19, Jan. 2023.

\bibitem{hece}
R. He \emph{et al.}, ``Propagation channels of 5G millimeter-wave vehicle-to-vehicle communications: Recent advances and future challenges,'' \emph{IEEE Veh. Technol. Mag.}, vol. 15, no. 1, pp. 16--26, Mar. 2020.



\bibitem{weather-2}
 R. T. Tan, ``Visibility in bad weather from a single image,'' in  \emph{Proc.
IEEE Conf. Comput. Vis. Pattern Recognit. (CVPR)}, Anchorage, AK, USA, Jun. 2008, pp. 1--8.
\bibitem{weather-communication}
X. You \emph{et al.}, ``Towards 6G wireless communication networks: Vision,
enabling technologies, and new paradigm shifts,'' Sci. China Inf. Sci., vol.
64, no. 1, Jan. 2021.

\bibitem{CARLA}
A. Dosovitskiy, G. Ros, F. Codevilla, A. Lopez, and V. Koltun,
``CARLA: An open urban driving simulator,'' in \emph{Proc. 1st Annu. Conf. Robot Learn.}, 2017, pp. 1--16.

\bibitem{LGSVL}
G. Rong \emph{et al.}, ``LGSVL simulator: A high fidelity simulator for autonomous driving,'' in \emph{Proc. IEEE 23rd Int. Conf. Intell. Transp.
Syst. (ITSC)}, 2020, pp. 1–6

\bibitem{DeepGTAV}
A. Ruano. Deepgtav code. [Online]. Available: https://github.com/aitorzip/DeepGTAV




\bibitem{sensory-1}
P. Sun \emph{et al.}, ``Scalability in perception for autonomous driving: Waymo Open Dataset," in \emph{Proc. IEEE/CVF Conf. Comput. Vis.
Pattern Recognit. (CVPR)}, Seattle, WA, USA, Jun. 2020, pp. 2443--2451.

\bibitem{sensory-2}
T. Renzler, M. Stolz, M. Schratter, and D. Watzenig, ``Increased accuracy for fast moving LiDARS: Correction of distorted point
clouds,'' in \emph{Proc. IEEE Int. Instrum. Meas. Technol. Conf. (I2MTC)},  Dubrovnik, Croatia,
May 2020, pp. 1--6.




\bibitem{Huang1}
 Z. Huang and X. Cheng, ``A general 3D space-time-frequency non-stationary model for 6G channels,'' \emph{IEEE Trans. Wireless Commun.},
vol. 20, no. 1, pp. 535--548, Jan. 2021.



\bibitem{weather-22}
S. G. Narasimhan and S. K. Nayar, ``Contrast restoration of weather degraded images,'' \emph{IEEE Trans. Pattern Anal. Mach. Intell.}, vol. 25, no. 6, pp. 713--724, Jun. 2003.



\bibitem{Huangj}
J. Huang \emph{et al.}, ``Multi-frequency mmWave massive MIMO channel measurements and characterization for 5G wireless communication systems," \emph{{IEEE} J. Select. Areas Commun.}, vol. 35, no. 7, pp. 1591--1605, Jul. 2017.

\bibitem{stuber}
G. L. Stüber, \emph{Principles of Mobile Communications}, 2nd ed. Norwell,
MA, USA: Kluwer, 2011.

\bibitem{TCOMmy}
Z. Huang, X. Cheng, and X. Yin, ``A general 3D non-stationary 6G
channel model with time-space consistency,'' \emph{IEEE Trans. Commun.},
vol. 70, no. 5, pp. 3436--3450, May 2022.

\bibitem{comst-weather}
S. A. Busari, K. M. S. Huq, S. Mumtaz, L. Dai, and J. Rodriguez, ``Millimeter-wave massive MIMO communication for future wireless systems: A survey," \emph{IEEE Commun. Surveys Tutor.}, vol. 20, no. 2, pp. 836--869, Secondquarter 2018.


	
\bibitem{sijiang} 
 X. Zheng \emph{et al.}, ``Confidence evaluation for machine learning schemes in vehicular sensor networks," \emph{IEEE Trans. Wireless Commun.}, vol. 22, no. 4, pp. 2833--2846, Apr. 2023.
	\end{thebibliography}
\end{document}